# Comorbidity Patterns and Temporal Associations of Multiple Long-Term Conditions in Adults with Intellectual Disability: an observational study in the United Kingdom

Georgina Cosma, Emeka Abakasanga, Rania Kousovista, Sharmin Shabnam, Navjot Kaur, Ashley Akbari, Reza Kiani, Francesco Zaccardi, Gyuchan Thomas Jun, Satheesh Gangadharan

Professor Georgina Cosma PhD, School of Science, Loughborough University, Loughborough, UK.
Dr Emeka Abakasanga PhD, School of Science, Loughborough University, Loughborough, UK.
Dr Rania Kousovista PhD, School of Science, Loughborough University, Loughborough, UK.
Dr Sharmin Shabnam PhD, Leicester Real World Evidence Unit, Leicester General Hospital, UK.
Dr Navjot Kaur, Leicester Real World Evidence Unit, Leicester General Hospital, UK.
Professor Ashley Akbari PhD, Faculty of Medicine, Health and Life Science, Swansea University, Swansea, UK.
Dr Reza Kiani MD, PhD, FRCPsych, Consultant Neuropsychiatrist and Honorary Associate Professor, Leicestershire Partnership NHS Trust & University of Leicester, UK.
Dr Francesco Zaccardi PhD, Leicester Real World Evidence Unit, Leicester General Hospital, UK.
Professor Gyuchan Thomas Jun PhD, School of Design and Creative Arts, Loughborough University, Loughborough, UK.
Professor Satheesh Gangadharan, Consultant Psychiatrist, Leicestershire Partnership NHS Trust, UK.

*Corresponding author: Professor G Cosma PhD; School of Science, Loughborough University, Loughborough, UK. Email: g.cosma@lboro.ac.uk. Phone: +44 (0)1509 564 158

**Objectives:** To examine the distribution, temporal associations, and age/sex-specific patterns of multiple long-term conditions (MLTCs) in adults with intellectual disability (ID).
**Study Design:** Observational study using longitudinal healthcare data.
**Methods:** Analysis of 18144 adults with ID (10168 males and 7976 females) identified in the Clinical Practice Research Datalink, linked to Hospital Episode Statistics Admitted Patient Care and Outpatient data (2000-2021). We used temporal analysis to establish directional associations among 40 long-term conditions, stratified by sex and age groups (under 45, 45-64, 65 and over).
**Results:** The high prevalence of enduring mental illness across all age groups is an important finding unique to this population. In males, mental illness occurred along with upper gastrointestinal conditions (specifically reflux disorders), while in females, mental illness presented alongside reflux disorders, chronic pain, and endocrine conditions such as thyroid problems. Among young males with intellectual disability, the combination of cerebral palsy with dysphagia, epilepsy, chronic constipation, and chronic pneumonia represents a distinctive pattern. In those aged 45-64, we observed early onset of lifestyle conditions like diabetes and hypertension, though notably these conditions co-occurred with mental illness and anaemia at rates exceeding those in the general population. The health conditions in those aged 65 and over largely mirrored those seen in the general aging population.
**Conclusions:** Our findings highlight the complex patterns of MLTCs in this population, revealing sex-specific associations across age groups, and identified temporal associations, thus providing insights into disease progression, which can inform targeted prevention strategies and interventions to prevent premature mortality.

## 1. Introduction

Multiple long-term conditions (MLTCs), defined by the National Institute of Health Research (NIHR) as the presence of two or more chronic health conditions (either physical or mental)[1], pose significant challenges for healthcare systems in Western countries, particularly for vulnerable populations such as individuals with intellectual disability[1,2]. This group often experiences earlier onset and higher prevalence of chronic conditions compared to the general population, leading to complex healthcare needs and reduced life expectancy[3–5]. Despite the growing recognition of this issue, the patterns of co-occurrence and progression of MLTCs in adults with intellectual disability remain poorly understood[4,6].
The complexity of managing MLTCs is compounded by the unique healthcare needs of individuals with intellectual disability[7]. These individuals often face barriers to healthcare access, difficulties in communication, and challenges in self-management of chronic conditions[8,9]. Furthermore, the interplay between intellectual disability and various chronic conditions can lead to atypical presentations and complications, making diagnosis and treatment more challenging. Previous research has highlighted the high prevalence of certain conditions in this population, such as epilepsy[10,11], mental illness[11,12], and chronic respiratory diseases[11,12]. However, there is a lack of comprehensive studies examining the co-occurrences of MLTCs in adults with intellectual disability[4]. Moreover, sex differences in the prevalence and progression of MLTCs in this population have not been thoroughly investigated[13].

Understanding the patterns of co-occurrence and progression of MLTCs is crucial for developing effective healthcare strategies[14]. It can inform clinical decision-making, guide preventive interventions and help in the design of integrated care pathways tailored to the needs of individuals with intellectual disability. Additionally, identifying sex and age-specific patterns can contribute to more personalised and effective healthcare approaches with the aim of reducing premature mortality in this population. In this study, we aimed to address these knowledge gaps by examining the distribution and associations of MLTCs in adults with intellectual disability, with a focus on sex and age differences.

## 1. Methods

### Study design and participants

Our study examined long-term condition (LTC) diagnosis records of individuals in the Clinical Practice Research Datalink (CPRD) identified as having an intellectual disability and aged 18 years or older between 1st January 2000 and 31st December 2021. We utilised primary care general practice (GP) records from the CPRD GOLD database alongside secondary care hospitalisation records from the Hospital Episode Statistics (HES) Admitted Patient Care (HES APC) and HES Outpatient (HES OP) data, focusing on a list of common LTCs relevant to the study cohort. This list was developed through a literature review followed by expert consensus from a professional advisory panel (PAP)[4,15].

After identifying the cohort, patients with a diagnosis of at least one of the selected LTCs from their birth until the study's end date were further extracted and stratified by sex, resulting in 10168 males and 7976 females. For each patient, the age at their first diagnosis of each condition (if applicable) was recorded for the analysis in this study. Table 1 presents the patient demographics, and Supp. Table S.0 presents the list of LTCs. Data collected at GPs were recorded using Read codes v2, whilst hospital data used International Classification of Diseases version 10 (ICD-10) codes. The complete list of Read v2 and ICD-10 codes for each condition, along with details on how chronicity was defined for each condition and how similar conditions were merged, is available[16]. The PAP effectively assessed the suitability of all the codes, ensuring that each condition was accurately represented.

It is important to note that in this study, intellectual disability (ID) is not classified as a long-term condition or disease. Therefore, ID is distinct from the mental illness condition. The mental illness condition excluded any codes on ID or behavioural disorders but included a set of clinical codes related to recurrent depression, bipolar disorder, chronic anxiety, and schizophrenia. Behavioural disorder was excluded due to inconsistency in the way it is coded making the estimation of prevalence challenging. Autism was not included as a LTC in this study as all individuals we studied have ID and a substantial proportion is expected to have autism. This study has been approved by the CPRD Independent Scientific Advisory Committee (protocol number: 22_001840).

### Methods for modelling condition progression

We designed an approach to highlight both strong individual associations and overall progression patterns in the dataset, allowing for a comprehensive interpretation of complex associations between conditions across different age groups. We initially stratified the data by sex (male, female) and then by age into three groups: below 45, 45-64, and 65 and over[17,18]. For each sex-age group combination, we created separate datasets containing the diagnosis dates for each condition. This stratification allows for sex- and age- specific analysis of condition progression. We employed Fisher's exact test to evaluate the statistical associations between pairs of conditions within each age group. A $p$-value<0.05 indicates a statistically significant association between conditions. To quantify the strength of associations between condition pairs, we calculated the odds ratio (OR) and its corresponding 95% confidence interval (CI).

Given the large number of condition pairs evaluated, we applied the False Discovery Rate (FDR) correction to mitigate the risk of false positive findings. We focused on associations with an adjusted p-value<0.05 after FDR correction. We examined the temporal associations between conditions by comparing their dates of occurrence when both were present in a patient to establish which condition typically preceded the other. We also computed the directional frequency and percentage to provide a comprehensive view of the associations between conditions.

We visualised chronic condition progression using directed graphs for females and males across the three age groups (<45, 45-64, 65+) (Figures 2-3)). In these graphs, nodes represent conditions, whilst edges depict progression between them. The edge direction indicates which condition typically precedes the other, based on the most common (i.e. frequent) temporal sequence of diagnoses for each pair. Edge thickness represents pair frequency, with thicker edges showing more common associations. Edge labels display ORs and median years between diagnoses. The legend associates arrow thickness with pair frequencies and data percentiles. For example, "144≤Frequency<205 (20%-40%)" indicates condition pairs occurring 144-204 times, falling within the 20th-40th percentile of observed frequencies. We included only statistically significant associations after FDR correction, applying a minimum prevalence threshold (adjusted for each age group) and requiring at least 100 patients per pair. Graph captions present group characteristics (sex, age group, total patients with diagnoses in this group) and visualised results (odds ratio range, observed minimum prevalence, number of condition pairs shown, total sum of odds ratios). Sensitivity analysis was performed to determine appropriate thresholds for including condition pairs in the network visualisations, ensuring the resulting graphs remained interpretable while capturing the most relevant clinical relationships. The results of the sensitivity analysis are shown in Supp. Table S.8.

## 2. Results

### Age and prevalence of chronic conditions

Table 1 shows the patient demographics, and a comparison of age statistics (age at the end of study date) and patient counts for chronic conditions across males and females is presented in Supp. Table S.1. Key observations from the data reveal that epilepsy is the most prevalent condition in males, affecting 35.1% of males (median age 18 years, IQR: 6-38) and 34.7% of females (median age 19 years, IQR: 6-40). This is higher than the prevalence that has already been reported by other researchers[19,20]. Long standing mental illness is the most common condition in females (39.5%, median age 35 years, IQR: 25-49) and the second most common in males (35.1%, median age 33 years, IQR: 23-47). Chronic airway diseases also show a high prevalence in both genders, affecting 24.6% of males (median age 21 years, IQR: 6-45) and 26.4% of females (median age 30 years, IQR: 14-48). Reflux disorders are another significant health condition, impacting 23.8% of males (median age 39 years, IQR: 27-51) and 26.2% of females (median age 41 years, IQR: 28-54). Hypertension also affects a substantial portion of the population, with 20.4% of males (median age 51 years, IQR: 41-62) and 22.3% of females (median age 53 years, IQR: 43-62) experiencing this condition. Supp. Figures 1a-1b show the heatmaps of co-occurring condition pairs among females and males, respectively. The box plots in Supp. Figures 2a-2b show the distribution of ages at first diagnosis of condition across sexes.

| Characteristic | Male | Female |
|---|---|---|
| **# Patients (%)** | 10168 | 7976 |
| **Age** | | |
| < 20 | 946 | 407 |
| 20-29 | 1778 | 1241 |
| 30-39 | 1716 | 1409 |
| 40-49 | 1575 | 1292 |
| 50-59 | 1831 | 1530 |
| 60-69 | 1304 | 1121 |
| 70-79 | 768 | 656 |
| 80+ | 250 | 320 |
| **Ethnic groups** | | |
| South Asian | 206 | 179 |
| Black | 233 | 167 |
| Mixed/Other | 211 | 191 |
| Unknown | 852 | 531 |
| White | 8666 | 6908 |
| **IMD Quintiles** | | |
| 1 (Least Deprived) | 1234 | 954 |
| 2 | 1795 | 1353 |
| 3 | 2193 | 1620 |
| 4 | 2313 | 1853 |
| 5 (Most Deprived) | 2619 | 2189 |
| Unknown | 14 | 7 |

**Table 1:** Patient demographics

### Comorbidity patterns

Figures 1-2 depict condition progression networks for females and males across three age groups (<45, 45-64, 65+). We have analysed condition combinations, focusing on frequency and prevalence: the results are shown in Tables 2-3. In Tables 2-3, the first column is the number of conditions in the combination; min. pair frequency is the smallest frequency of any pair within the combination; and the last column shows the percentage of prevalence of the combination in the specific group. In the discussions that follow, only condition pairs with a frequency of 100 or more were included to focus on the most prevalent associations in each group, and precedence between conditions was only mentioned if a specific directionality occurred in more than 70% of the cases. The results of the analysis on the association between conditions stratified by sex and age group are presented in Supp. Tables S.2-S.7.

**Females under 45 (n=6397).** Firstly, the strongest association is observed between cerebral palsy and epilepsy with an OR of 5.29 (CI: 4.53-6.18). Cerebral palsy also shows significant associations with dysphagia (OR: 5.35, CI: 4.30-6.66) and visual impairment (OR: 4.17, CI: 3.32-5.24) which are all in line with the published literature[21]. Chronic airway diseases precede reflux disorders in 70.15% of cases and chronic pain conditions in 70.45% of cases and have significant associations with reflux disorders (OR: 2.60, CI: 2.27-2.98), insomnia (OR: 2.80, CI: 2.39-3.27), and chronic pain conditions (OR: 3.17, CI: 2.64-3.79). Diabetes and hypertension show a strong association with each other (OR: 9.73, CI: 7.76-12.19). Chronic pain conditions show significant associations with reflux disorders (OR: 3.63, CI: 3.02-4.36). Mental illness demonstrates associations with several conditions, including insomnia (OR: 3.09, CI: 2.66-3.58), chronic pain conditions (OR: 3.07, CI: 2.58-3.66), and neuropathic pain (OR: 3.14, CI: 2.46-4.02). Reflux disorders appear to have strong associations with various gastrointestinal conditions including chronic diarrhoea (OR: 5.34, CI: 4.10-6.96), dysphagia (OR: 4.39, CI: 3.57-5.39), and chronic constipation (OR: 3.99, CI: 3.20-4.97).

**Females between 45 and 64 (n=3494).** The strongest association in this age group of females is observed between diabetes and hypertension with an OR of 9.39 (CI: 7.74-11.40) which is also the highest occurrence, i.e. 6.75% of this group. Hypertension shows strong associations with reflux disorders (OR: 4.44, CI: 3.70-5.32), chronic kidney disease (CKD) (OR: 4.89, CI: 4.04-5.91), and chronic arthritis (OR: 4.02, CI: 3.34-4.85). In addition to hypertension, CKD appears to have significant associations with anaemia (OR: 5.70, CI: 4.63-7.03) and diabetes (OR: 3.94, CI: 3.11-4.99). Diabetes precedes CKD in 71.57% of cases, which is one of the highest precedence percentages in the group. Mental illness demonstrates associations with several conditions, including reflux disorders (OR: 4.72, CI: 3.91-5.70), menopausal and perimenopausal disorders (OR: 3.78, CI: 3.12-4.59), and insomnia (OR: 4.84, CI: 3.85-6.09). Dementia shows a particularly strong association with epilepsy (OR: 16.83, CI: 13.27-21.34). Chronic arthritis demonstrates significant associations with reflux disorders (OR: 5.32, CI: 4.37-6.48), menopausal and perimenopausal disorders (OR: 4.34, CI: 3.55-5.32), and chronic airway diseases (OR: 5.72, CI: 4.59-7.13). Coronary heart disease shows a strong association with hypertension (OR: 11.49, CI: 8.58-15.38). Dysphagia demonstrates a significant association with reflux disorders (OR: 6.81, CI: 5.45-8.52).

**Females 65 and over (n=1292).** The strongest association in this group is observed between CKD and heart failure with an OR of 32.76 (CI: 24.35-44.09), and this association occurs in 9.29% of the group. CKD shows significant associations with hypertension (OR: 20, CI: 15.71-25.47), anaemia (OR: 23.80, CI: 18.44-30.72), arrhythmias (OR: 22.77, CI: 17.64-29.39), and dementia (OR: 19.75, CI: 15.30-25.50). The association between CKD and hypertension occurs in 11.84% of the group, which is the highest frequency in this group. Arrhythmias demonstrate significant associations with CKD, hypertension (OR: 18.91, CI: 14.50-24.65) and heart failure (OR: 39.96, CI: 29.42-54.28). Hypertension has strong associations with various conditions including chronic arthritis (OR: 20.71, CI: 15.79-27.14), anaemia (OR: 15.93, CI: 12.19-20.82), and dementia (OR: 15.63, CI: 11.96-20.44). Hypertension precedes dementia in 75.24% of cases. Chronic arthritis shows a significant association with CKD (OR: 14.34, CI: 10.99-18.71). Anaemia demonstrates strong associations with arrhythmias (OR: 19.16, CI: 14.52-25.27) in addition to its associations with CKD and hypertension.

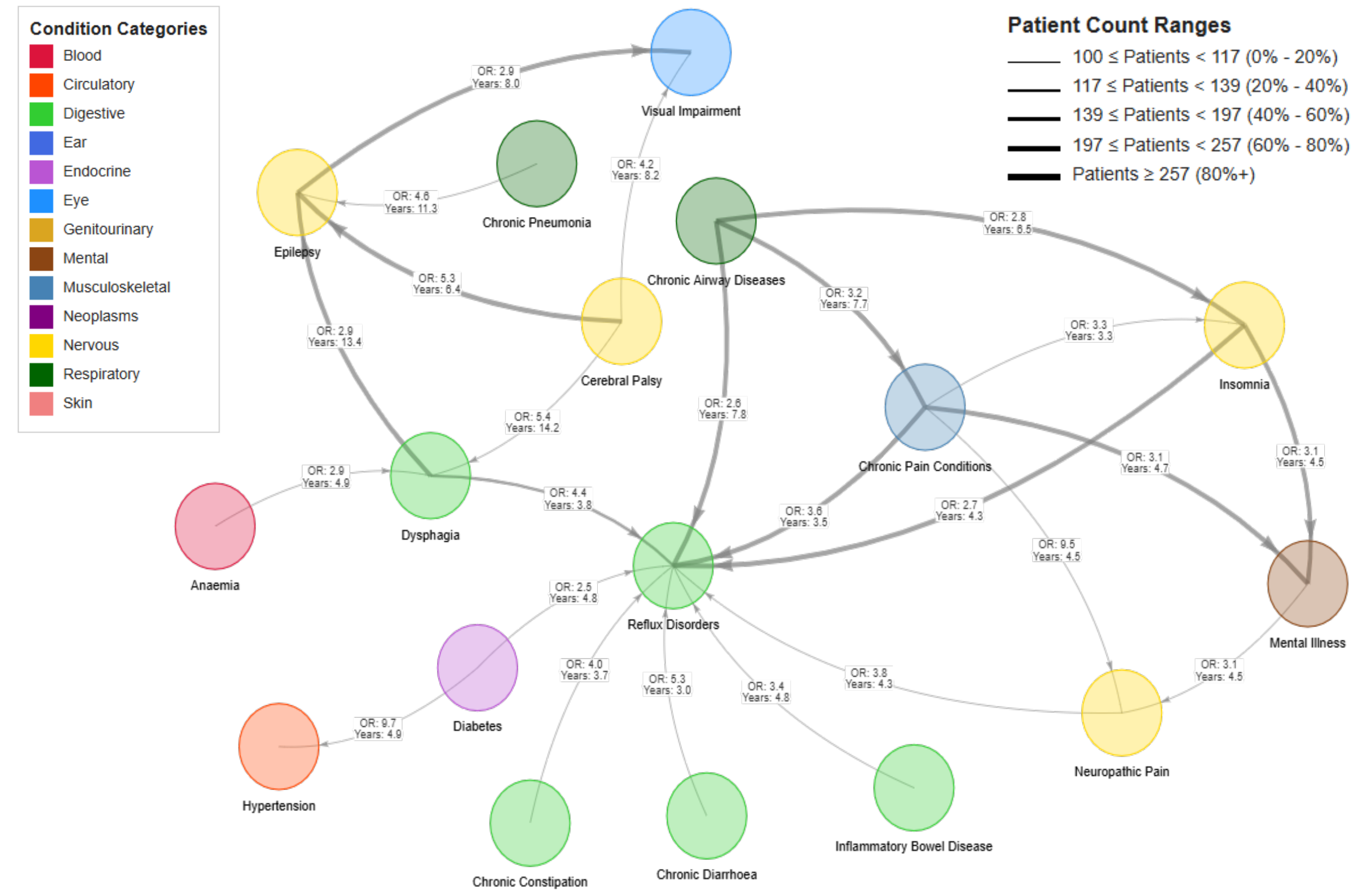

**(a) Sex:** Female. **Age group:** <45. **Total patients with diagnoses in this group:** 6397. **Odds ratio range:** [2.48-9.73]. **Observed minimum prevalence:** 1.56% (99 patients). **Number of condition pairs shown:** 24. **Total sum of odds ratios:** 98.20.

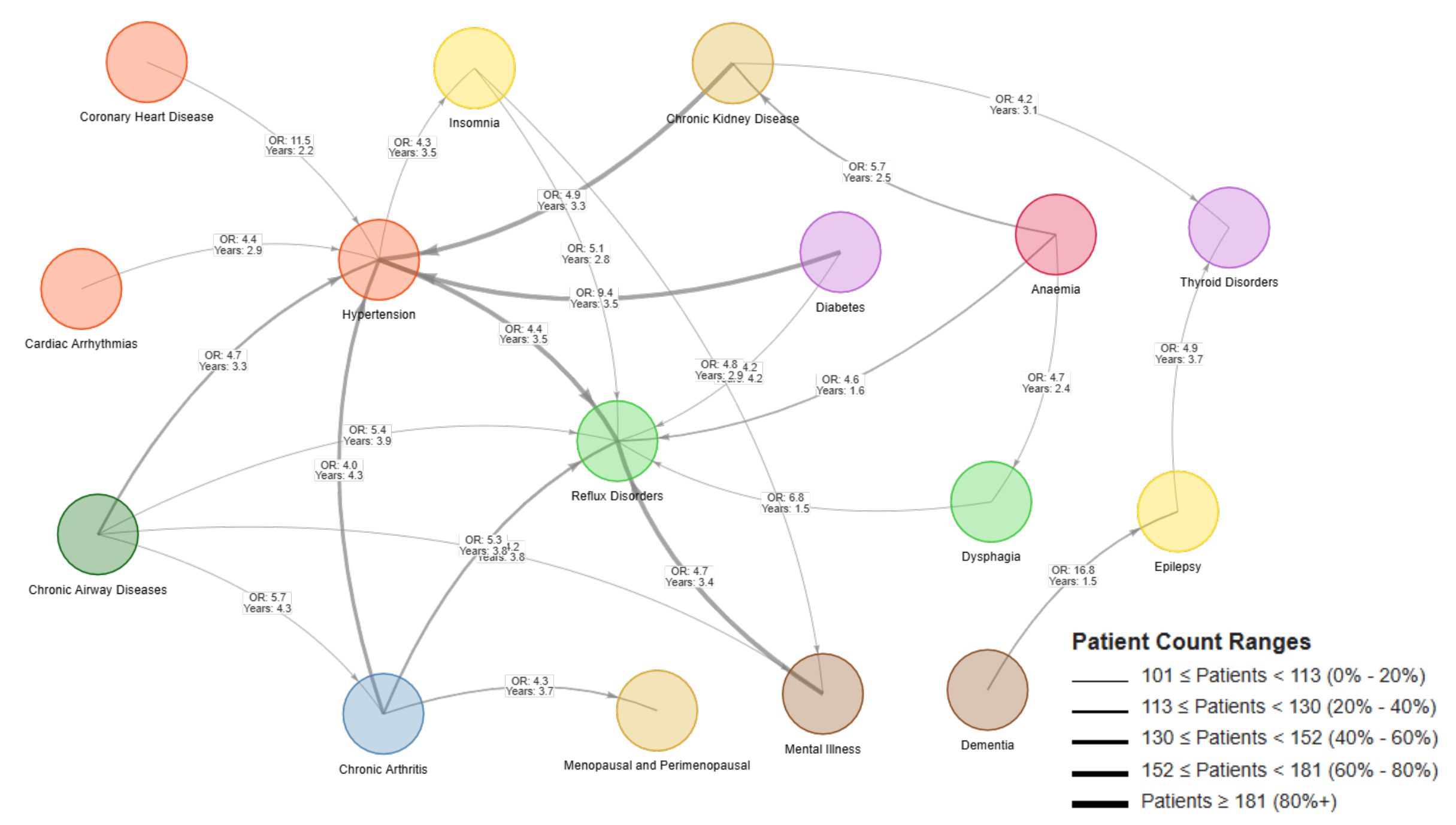

**(b) Sex:** Female. **Age group:** 45-64. **Total patients with diagnoses in this group:** 3494. **Odds ratio range:** [4.02 - 16.83]. **Observed minimum prevalence:** 2.89% (101 patients). **Number of condition pairs shown:** 24. **Total sum of odds ratios:** 139.33.

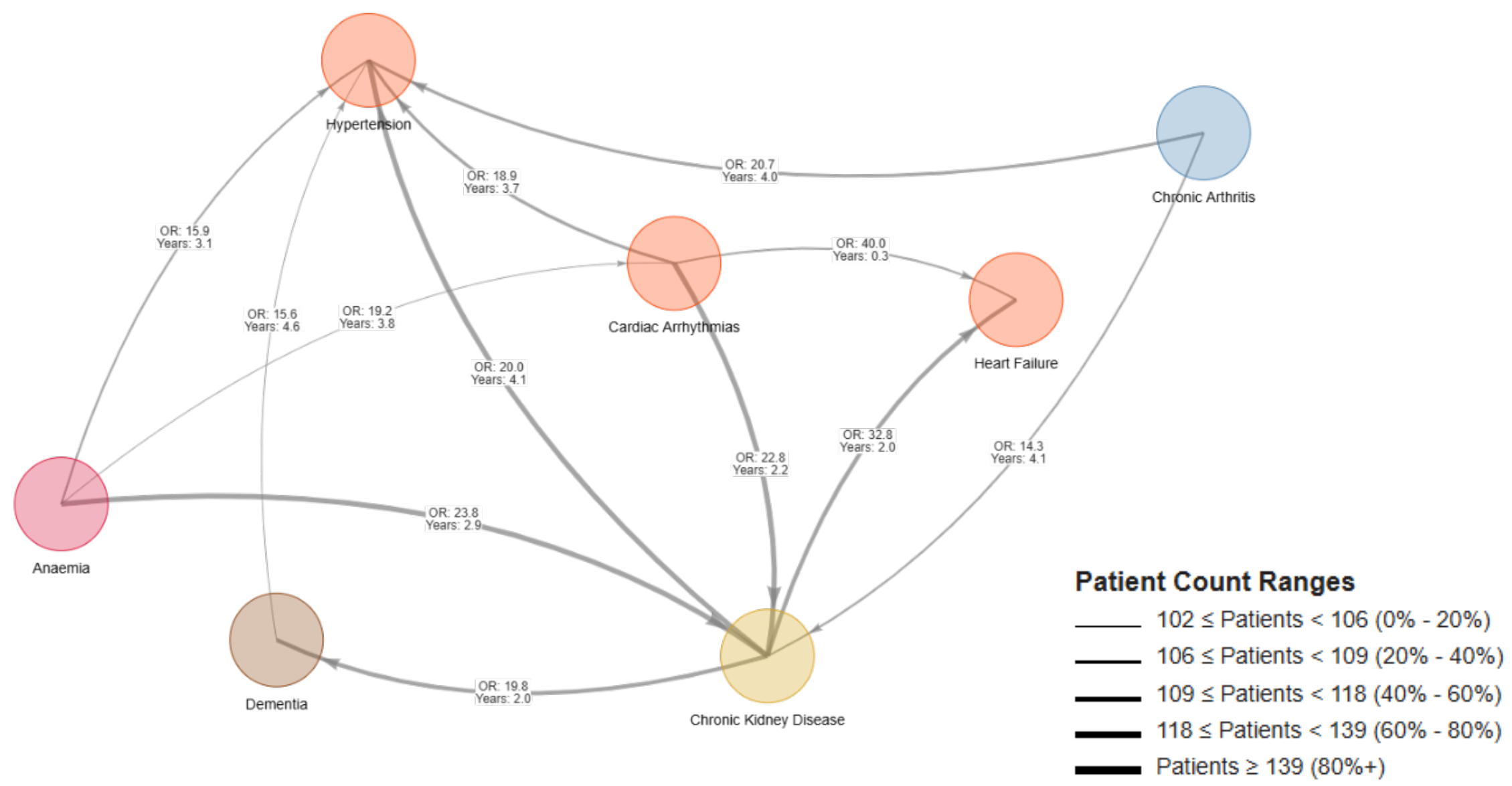

**(3) Sex:** Female**. Age group:** 65+**. Total patients with diagnoses in this group:** 1292**. Odds ratio range:** [14.34-39.96]**. Observed minimum prevalence:** 7.89% (101 patients)**. Number of condition pairs shown:** 12**. Total sum of odds ratios:** 263.73

**Figure 1:** Progression networks for females across age groups. The threshold OR values were chosen based on the outcomes of the sensitivity analysis (see Supp. Table 8). Only pairs with frequency >=100 are shown.

| Females | | | |
|---|---|---|---|
| **Combination** | **No. of conditions** | **Min. pair frequency** | **Prevalence of combination %** |
| | | | |
| **Females under 45 years** | | | |
| Chronic Airway Diseases + Insomnia + Reflux Disorders | 3 | 239 | 3.74 |
| Chronic Airway Diseases + Chronic Pain Conditions + Reflux Disorders | 3 | 203 | 3.17 |
| Cerebral Palsy + Dysphagia + Epilepsy | 3 | 139 | 2.17 |
| Chronic Airway Diseases + Chronic Pain Conditions + Insomnia | 3 | 136 | 2.13 |
| Chronic Pain Conditions + Insomnia + Mental Illness | 3 | 136 | 2.13 |
| Chronic Pain Conditions + Insomnia + Reflux Disorders | 3 | 136 | 2.13 |
| Chronic Airway Diseases + Chronic Pain Conditions + Insomnia + Reflux Disorders | 4 | 136 | 2.13 |
| Anaemia + Dysphagia + Reflux Disorders | 3 | 121 | 1.89 |
| Cerebral Palsy + Epilepsy + Visual Impairment | 3 | 118 | 1.84 |
| Chronic Pain Conditions + Mental Illness + Neuropathic Pain | 3 | 100 | 1.56 |
| Chronic Pain Conditions + Neuropathic Pain + Reflux Disorders | 3 | 100 | 1.56 |
| | | | |
| **Females 45-64 years** | | | |
| Hypertension + Mental Illness + Reflux Disorders | 3 | 191 | 5.47 |
| Chronic Arthritis + Hypertension + Reflux Disorders | 3 | 175 | 5.01 |
| Hypertension + Menopausal and Perimenopausal + Mental Illness | 3 | 167 | 4.78 |
| Anaemia + Hypertension + Reflux Disorders | 3 | 154 | 4.41 |
| Chronic Arthritis + Hypertension + Menopausal and Perimenopausal | 3 | 154 | 4.41 |
| Anaemia + chronic kidney disease + Hypertension | 3 | 152 | 4.35 |
| Chronic Arthritis + Hypertension + Mental Illness | 3 | 152 | 4.35 |
| Chronic Arthritis + Menopausal and Perimenopausal + Mental Illness | 3 | 152 | 4.35 |
| Chronic Arthritis + Menopausal and Perimenopausal + Reflux Disorders | 3 | 152 | 4.35 |
| Chronic Arthritis + Mental Illness + Reflux Disorders | 3 | 152 | 4.35 |
| Hypertension + Menopausal and Perimenopausal + Reflux Disorders | 3 | 152 | 4.35 |
| Menopausal and Perimenopausal + Mental Illness + Reflux Disorders | 3 | 152 | 4.35 |
| Chronic Arthritis + Hypertension + Menopausal and Perimenopausal + Mental Illness | 4 | 152 | 4.35 |
| Chronic Arthritis + Hypertension + Menopausal and Perimenopausal + Reflux Disorders | 4 | 152 | 4.35 |
| Chronic Arthritis + Hypertension + Mental Illness + Reflux Disorders | 4 | 152 | 4.35 |
| Chronic Arthritis + Menopausal and Perimenopausal + Mental Illness + Reflux Disorders | 4 | 152 | 4.35 |
| Hypertension + Menopausal and Perimenopausal + Mental Illness + Reflux Disorders | 4 | 152 | 4.35 |
| Chronic Arthritis + Hypertension + Menopausal and Perimenopausal + Mental Illness + Reflux Disorders | 5 | 152 | 4.35 |
| | | | |
| **Females 65+ years** | | | |
| Cardiac Arrhythmias + chronic kidney disease + Hypertension | 3 | 115 | 8.9 |
| Chronic Arthritis + chronic kidney disease + Hypertension | 3 | 107 | 8.28 |
| Anaemia + chronic kidney disease + Hypertension | 3 | 106 | 8.2 |
| Cardiac Arrhythmias + chronic kidney disease + Heart Failure | 3 | 106 | 8.2 |
| Chronic kidney disease + Dementia + Hypertension | 3 | 105 | 8.13 |
| Anaemia + Cardiac Arrhythmias + Chronic Kidney Disease | 3 | 102 | 7.89 |
| Anaemia + Cardiac Arrhythmias + Hypertension | 3 | 102 | 7.89 |
| Anaemia + Cardiac Arrhythmias + chronic kidney disease + Hypertension | 4 | 102 | 7.89 |

**Table 2:** Combinations of conditions with minimum pair frequency of >=100 for females across age groups. The conditions within each combination are in no order. For females 45-64 years a prevalence threshold was applied (>=4%) due to the large number of combinations.

**Males under 45 (n=8296).** The most frequent association is between cerebral palsy and epilepsy (OR: 4.27, CI: 3.74-4.87). Cerebral palsy also shows strong associations with dysphagia (OR: 6.26, CI: 5.18-7.57) and visual impairment (OR: 4.44, CI: 3.66-5.40). Dysphagia is frequently associated with epilepsy (OR: 2.42, CI: 2.03-2.88) and reflux disorders (OR: 4.47, CI: 3.72-5.36). Diabetes and hypertension show the strongest association (OR: 8.51, CI: 6.92-10.48). Other significant associations include anaemia with reflux disorders (OR: 3.13, CI: 2.61-3.74), chronic constipation with epilepsy (OR: 2.51, CI: 2.02-3.13), and inflammatory bowel disease with reflux disorders (OR: 3.13, CI: 2.52-3.89). Epilepsy precedes chronic constipation in 75.61% of cases. Chronic pneumonia and epilepsy also show a significant association (OR: 4.10, CI: 3.12-5.39), with epilepsy preceding chronic pneumonia in 72.79% of cases.

**Males between 45 and 64 (n= 3969).** The most frequent association is between diabetes and hypertension (OR: 9.73, CI: 8.10-11.68). Hypertension also shows strong associations with mental illness (OR: 4.03, CI: 3.39-4.78) and reflux disorders (OR: 4.55, CI: 3.82-5.41). CKD has significant associations with hypertension (OR: 5.36, CI: 4.46-6.43), anaemia (OR: 9.12, CI: 7.45-11.16), and diabetes (OR: 5.71, CI: 4.59-7.10). Coronary heart disease shows strong associations with hypertension (OR: 9.40, CI: 7.56-11.68) and cardiac arrhythmias (OR: 13.56, CI: 10.61-17.32). Anaemia is frequently associated with hypertension (OR: 4.53, CI: 3.75-5.48) and reflux disorders (OR: 7.24, CI: 5.95-8.81). Other notable associations include dementia with epilepsy (OR: 15.48, CI: 12.35-19.40) and dysphagia (OR: 9.29, CI: 7.31-11.82), as well as cardiac arrhythmias with heart failure (OR: 21.20, CI: 16.05-28.01). The strongest association is between coronary heart disease and heart failure (OR: 27.51, CI: 20.67-36.59).

**Males 65 and over (n=1413).** CKD appears to have strong associations with several conditions in this age group, the strongest and the most frequent of which is observed between CKD and hypertension (in 12.74%) with an OR of 26.19 (CI: 20.82-32.95). CKD also shows significant associations with arrhythmias (OR: 27.06, CI: 21.47-34.10), anaemia (OR: 26.34, CI: 20.51-33.84), heart failure (OR: 33.41, CI: 25.04-44.57), dementia (OR: 14.52, CI: 11.19-18.84), and diabetes (OR: 23.06, CI: 17.28-30.77). Arrhythmias demonstrate strong associations with CKD, hypertension (OR: 19.62, CI: 15.32-25.13), and heart failure (OR: 39.99, CI: 29.81-53.65). Anaemia shows significant associations with CKD and arrhythmias (OR: 21.26, CI: 16.31-27.71). The association between cardiac arrhythmias and CKD is the second most frequent, occurring in 12.67% of the group, followed by anaemia and CKD at 10.47%.

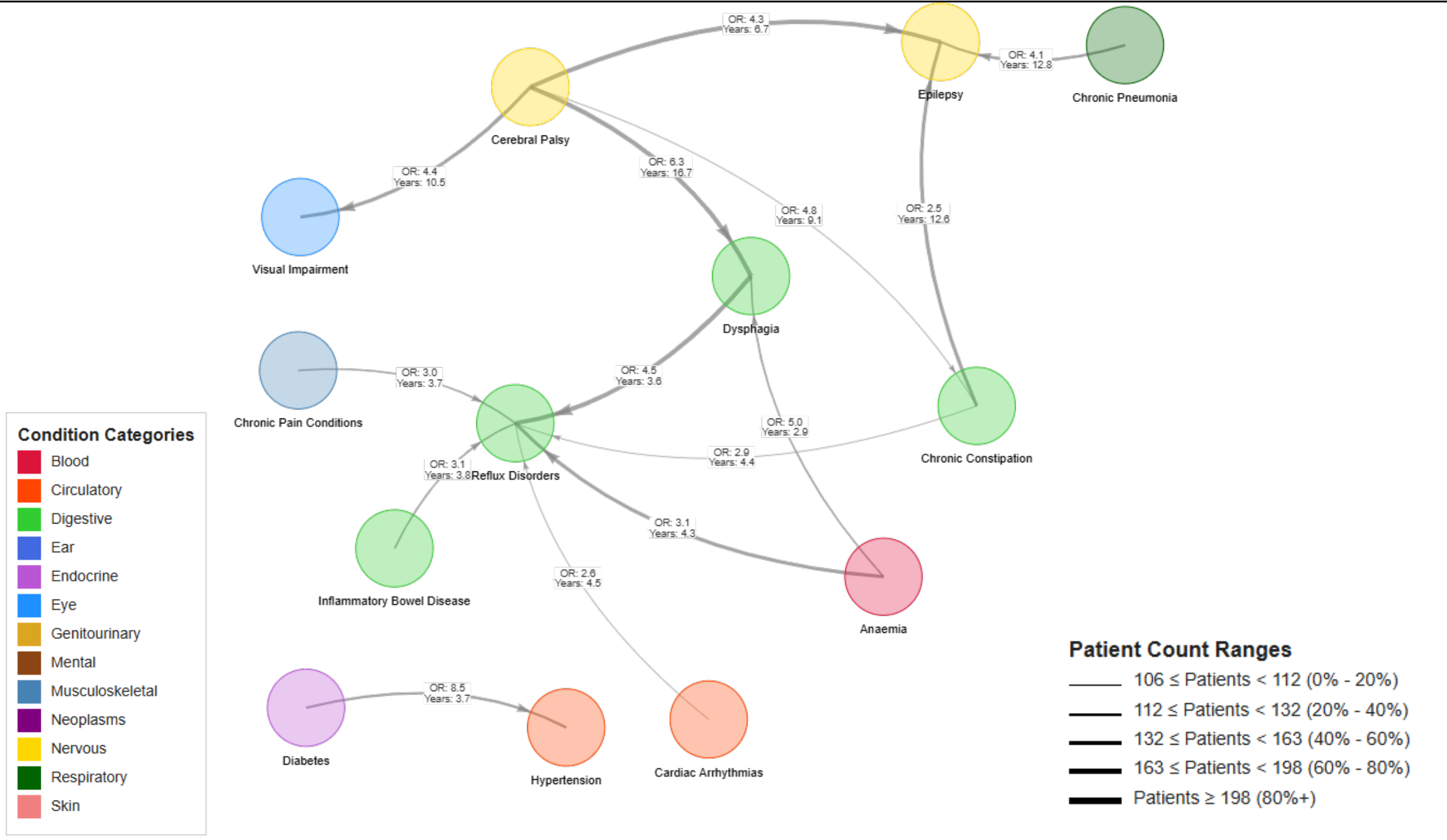

**(a) Sex: Male. Age group:** <45**. Total patients with diagnoses in this group:** 8296**. Odds ratio range:** [2.51 - 8.51]**. Observed minimum prevalence:** 1.28% (105 patients)**. Number of condition pairs shown:** 14**. Total sum of odds ratios:** 59.08

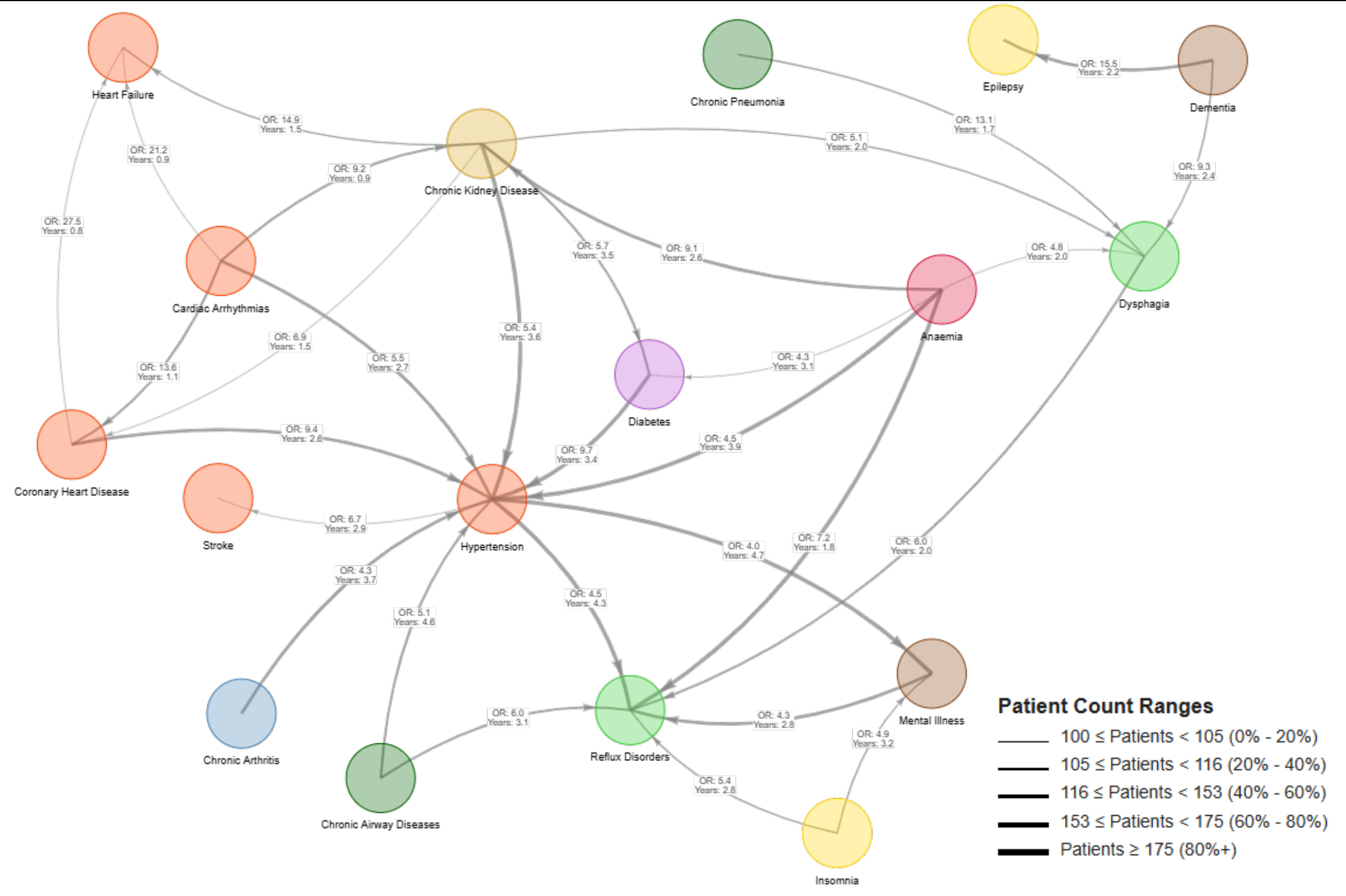

**(b) Sex: Male. Age group:** 45-64. **Total patients with diagnoses in this group:** 3969. **Odds ratio range:** [4.03-27.51]. **Observed minimum prevalence:** 2.52% (100 patients). **Number of condition pairs shown:** 30. **Total sum of odds ratios:** 253.37

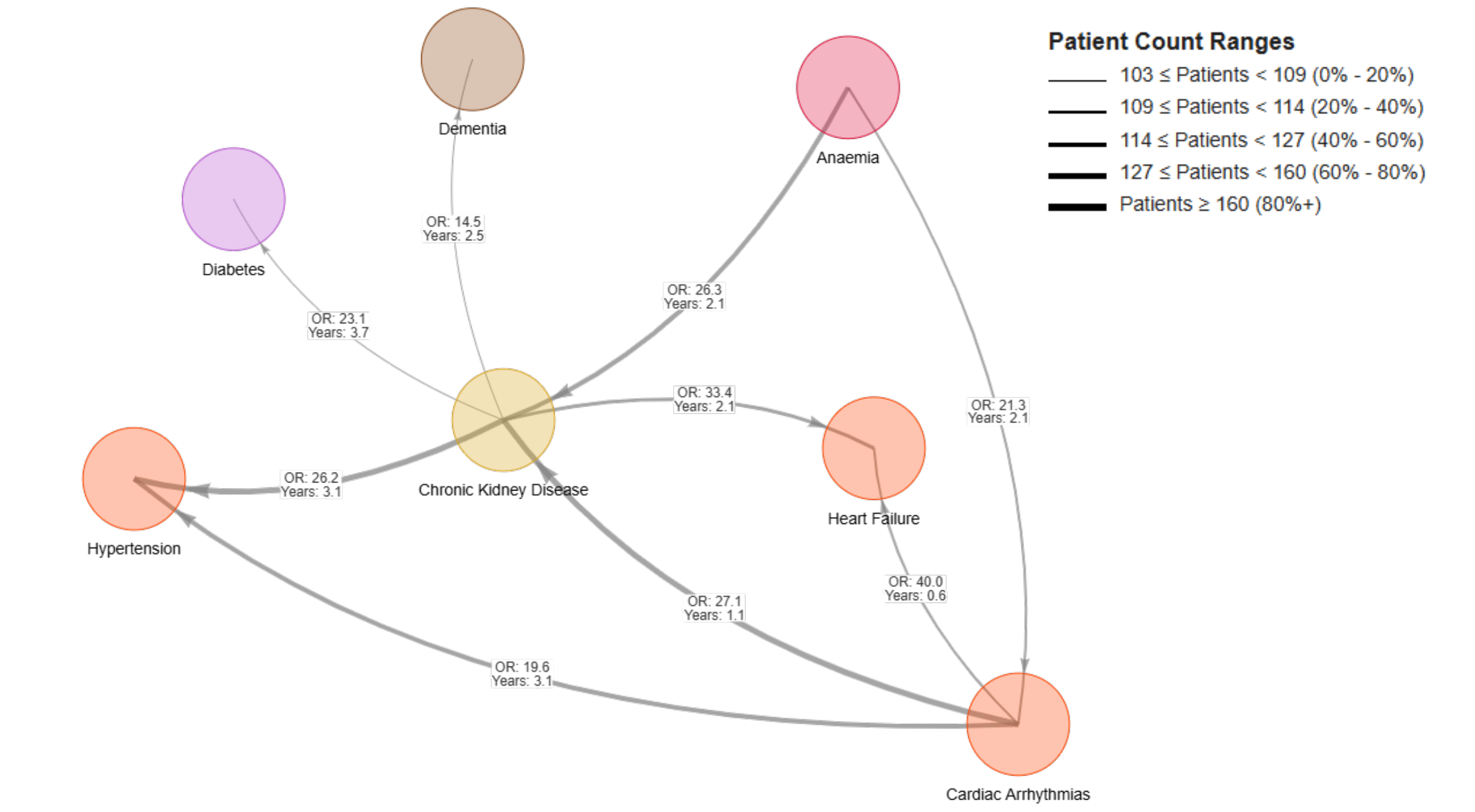

**(c) Sex:** Male. Age group: 65+. **Total patients with diagnoses in this group:** 1413. **Odds ratio range:** [14.52-39.99]. **Observed minimum prevalence:** 7.29% (103 patients). **Number of condition pairs shown:** 9. **Total sum of odds ratios:** 231.45

**Figure 2:** Progression networks for males across age groups. The threshold OR values were chosen based on the outcomes of the sensitivity analysis (see Supp. Table 8). Only pairs with frequency >=100 are shown.

| Males | | | |
|---|---|---|---|
| **Combinations** | **No. of conditions** | **Min. pair frequency** | **Prevalence of combination %** |
| **Males under 45** | | | |
| Cerebral Palsy + Dysphagia + Epilepsy | 3 | 200 | 2.41 |
| Anaemia + Dysphagia + Reflux Disorders | 3 | 112 | 1.35 |
| Cerebral Palsy + Chronic Constipation + Epilepsy | 3 | 111 | 1.34 |
| Cerebral Palsy + Chronic Pneumonia + Epilepsy | 3 | 94 | 1.13 |
| Cerebral Palsy + Chronic Pneumonia + Dysphagia | 3 | 88 | 1.06 |
| Chronic Pneumonia + Dysphagia + Epilepsy | 3 | 88 | 1.06 |
| Cerebral Palsy + Chronic Pneumonia + Dysphagia + Epilepsy | 4 | 88 | 1.06 |
| Chronic Pneumonia + Dysphagia + Reflux Disorders | 3 | 86 | 1.04 |
| Cerebral Palsy + Dysphagia + Visual Impairment | 3 | 85 | 1.02 |
| | | | |
| **Males 45-64 years** | **No. of conditions** | **Min. pair frequency** | **Prevalence of combination %** |
| Anaemia + Hypertension + Reflux Disorders | 3 | 177 | 4.46 |
| Anaemia + Chronic kidney disease + Hypertension | 3 | 175 | 4.41 |
| Hypertension + Mental Illness + Reflux Disorders | 3 | 170 | 4.28 |
| Cardiac Arrhythmias + Chronic kidney disease + Hypertension | 3 | 144 | 3.63 |
| Chronic Airway Diseases + Hypertension + Reflux Disorders | 3 | 128 | 3.22 |
| Chronic kidney disease + Diabetes + Hypertension | 3 | 127 | 3.2 |
| Chronic kidney disease + Hypertension + Mental Illness | 3 | 126 | 3.17 |
| Anaemia + Chronic kidney disease + Mental Illness | 3 | 123 | 3.1 |
| Anaemia + Chronic kidney disease + Reflux Disorders | 3 | 123 | 3.1 |
| Anaemia + Hypertension + Mental Illness | 3 | 123 | 3.1 |
| Anaemia + Mental Illness + Reflux Disorders | 3 | 123 | 3.1 |
| Chronic kidney disease + Hypertension + Reflux Disorders | 3 | 123 | 3.1 |
| Chronic kidney disease + Mental Illness + Reflux Disorders | 3 | 123 | 3.1 |
| Anaemia + Chronic kidney disease + Hypertension + Mental Illness | 4 | 123 | 3.1 |
| Anaemia + Chronic kidney disease + Hypertension + Reflux Disorders | 4 | 123 | 3.1 |
| Anaemia + Chronic kidney disease + Mental Illness + Reflux Disorders | 4 | 123 | 3.1 |
| Anaemia + Hypertension + Mental Illness + Reflux Disorders | 4 | 123 | 3.1 |
| Chronic kidney disease + Hypertension + Mental Illness + Reflux Disorders | 4 | 123 | 3.1 |
| Anaemia + Chronic kidney disease + Hypertension + Mental Illness + Reflux Disorders | 5 | 123 | 3.1 |
| | | | |
| **Males 65+ years** | **No. of conditions** | **Min. pair frequency** | **Prevalence of combination %** |
| Cardiac Arrhythmias + Chronic kidney disease + Hypertension | 3 | 128 | 9.06 |
| Cardiac Arrhythmias + Chronic kidney disease + Heart Failure | 3 | 112 | 7.93 |
| Anaemia + Cardiac Arrhythmias + Chronic Kidney Disease | 3 | 110 | 7.78 |
| Anaemia + Cardiac Arrhythmias + Hypertension | 3 | 97 | 6.86 |
| Anaemia + Chronic kidney disease + Hypertension | 3 | 97 | 6.86 |
| Cardiac Arrhythmias + Chronic kidney disease + Coronary Heart Disease | 3 | 97 | 6.86 |
| Anaemia + Cardiac Arrhythmias + Chronic kidney disease + Hypertension | 4 | 97 | 6.86 |
| Cardiac Arrhythmias + Coronary Heart Disease + Hypertension | 3 | 90 | 6.37 |
| Chronic Kidney Disease + Coronary Heart Disease + Hypertension | 3 | 90 | 6.37 |
| Cardiac Arrhythmias + Chronic Kidney Disease + Coronary Heart Disease + Hypertension | 4 | 90 | 6.37 |
| Chronic Kidney Disease + Heart Failure + Hypertension | 3 | 88 | 6.23 |
| Cardiac Arrhythmias + Heart Failure + Hypertension | 3 | 88 | 6.23 |
| Cardiac Arrhythmias + Chronic Kidney Disease + Heart Failure + Hypertension | 4 | 88 | 6.23 |
| Chronic Kidney Disease + Diabetes + Hypertension | 3 | 86 | 6.09 |
| Cardiac Arrhythmias + Chronic Kidney Disease + Dementia | 3 | 86 | 6.09 |
| Cardiac Arrhythmias + Chronic Kidney Disease + Dementia + Hypertension | 4 | 85 | 6.02 |
| Cardiac Arrhythmias + Dementia + Hypertension | 3 | 85 | 6.02 |
| Chronic Kidney Disease + Dementia + Hypertension | 3 | 85 | 6.02 |

**Table 3:** Combinations of conditions with minimum pair frequency of >=100 for males across age groups. The conditions within each combination are in no order. For males 45-64 years only the combinations with a prevalence of >=3% are shown, and for males over 65 the combinations with a prevalence of >=6% are shown.

## 3. Discussion

Previous investigations have examined the prevalence and patterns of multiple long-term conditions (MLTCs) in people with intellectual disability. A systematic scoping review found that across studies, the prevalence of MLTCs (defined as 2 or more conditions) ranged from 8-99% in this population[4,6]. This contrasts with the general population, where such conditions typically emerge around age 65 or older[22]. Within the literature, the most common conditions reported for the population of people with MLTCs and intellectual disability were epilepsy, visual impairment, and thyroid disorders[4]. Where reported, mental health issues commonly co-occurred with other conditions like obesity and epilepsy. Physical condition combinations included visual impairment with constipation, and epilepsy with constipation or visual impairment. A recent review highlighted a lack of studies examining sex differences and temporal relationships between conditions[4].

Within our study, in all age groups, the high prevalence of enduring mental illness (which does not include intellectual disability or behavioural disorder) is an important finding unique to this population, particularly among females. People with ID in our study had higher rates of several health conditions compared to the general population, particularly for epilepsy, mental illness, circulatory and respiratory conditions (see Supp. Table S.9). The high prevalence of enduring mental illness across all age groups is an important finding unique to this population. In males, mental illness occurred along with upper gastrointestinal conditions (specifically reflux disorders), while in females, mental illness presented alongside reflux disorders, chronic pain, and endocrine conditions such as thyroid problems. Among young males with intellectual disability, the combination of cerebral palsy with dysphagia, epilepsy, chronic constipation, and chronic pneumonia represents a distinctive pattern. Some of these conditions are noted in LeDeR[5] as conditions responsible for higher mortality in this population. In those aged 45-64, we observed early onset of lifestyle conditions like diabetes and hypertension, though notably these conditions co-occurred with mental illness and anaemia at rates exceeding those in the general population. The health conditions in those aged 65 and over largely mirrored those seen in the general aging population.

In conclusion, the high prevalence and early onset of MLTC in people with intellectual disability are consistently reported across studies, including ours. Our findings on the nature of concurrent conditions, sex differences, age-specific patterns, and temporal associations between conditions provide a more detailed and comprehensive understanding of MLTCs in this population. These disparities which have been reported in the general population[23,24] as well demonstrate the need for sex-specific screening for early identification and management of these conditions. Equally, the disease trajectory could be influenced by other biological, social, cultural and psychological factors[3,16,17,21,25–29] which have not been explored in this study. This evidence presented in our study can inform the development of targeted screening and prevention strategies, particularly for conditions identified as precursors to MLTCs. Healthcare providers should be aware of common condition combinations to improve early detection and integrated care approaches.

While our study offers important insights, there are limitations to consider. The UK focus may restrict direct applicability to countries with different healthcare systems or population characteristics. Furthermore, the study population comprises individuals who have accessed healthcare services, potentially underrepresenting those facing barriers to access. While our 22-year study period provides comprehensive longitudinal data, changes in diagnostic practices over this timeframe may have influenced the observed patterns.

Future research should focus on validating the identified progression patterns in diverse populations, investigating the underlying mechanisms of MLTC development in people with intellectual disability, and examining similar patterns across different global healthcare systems. This would further enhance our understanding of the healthcare needs of adults with intellectual disability across varied populations and contexts.

**Acknowledgments.** Data-driven machinE-learning aided stratification and management of multiple long-term COnditions in adults with intellectual disabilitiEs (DECODE) project (NIHR203981) is funded by the NIHR AI for Multiple Long-term Conditions (AIM) Programme. The views expressed are those of the author(s) and not necessarily those of the NIHR or the Department of Health and Social Care. This work uses data provided by patients and collected by the NHS as part of their care and support. We also want to acknowledge all data providers who make anonymised data available for research.

**Role of the funding source.** The funders of the study had no role in study design, data collection, data analysis, data interpretation, or writing of the report.

**Declaration of interests.** The authors declare no competing interests.

**Data sharing.** The individual-level case data cannot be shared publicly. The tables that were utilised to generate the figures shown in the paper are provided in the Supplementary Appendix.

**References**

[1] Dambha-Miller H, Cheema S, Saunders N, Simpson G. Multiple long-term conditions (MLTC) and the environment: a scoping review. Int J Environ Res Public Health 2022;19:11492.

[2] Valabhji J, Barron E, Pratt A, Hafezparast N, Dunbar-Rees R, Turner EB, et al. Prevalence of multiple long-term conditions (multimorbidity) in England: a whole population study of over 60 million people. J R Soc Med 2024;117:104–17.

[3] Emerson E, Hatton C. Health inequalities and people with intellectual disabilities. Cambridge University Press; 2014.

[4] Mann C, Jun GT, Tyrer F, Kiani R, Lewin G, Gangadharan SK. A scoping review of clusters of multiple long-term conditions in people with intellectual disabilities and factors impacting on outcomes for this patient group. Journal of Intellectual Disabilities 2023;27:1045–61.

[5] White A, Sheehan R, Ding J, Roberts C, Magill N, Keagan-Bull R, et al. Learning from Lives and Deaths - People with a learning disability and autistic people (LeDeR) report for 2022. King's College London 2023.

[6] Kinnear D, Morrison J, Allan L, Henderson A, Smiley E, Cooper S-A. Prevalence of physical conditions and multimorbidity in a cohort of adults with intellectual disabilities with and without Down syndrome: cross-sectional study. BMJ Open 2018;8:e018292.

[7] Heslop P, Blair PS, Fleming P, Hoghton M, Marriott A, Russ L. The Confidential Inquiry into premature deaths of people with intellectual disabilities in the UK: a population-based study. The Lancet 2014;383:889–95.

[8] Shea Bradley and Bailie J and DSH and FN and LN and BR. Access to general practice for people with intellectual disability in Australia: a systematic scoping review. BMC Primary Care 2022;23:306.

[9] Ali A, Scior K, Ratti V, Strydom A, King M, Hassiotis A. Discrimination and other barriers to accessing health care: perspectives of patients with mild and moderate intellectual disability and their carers. PLoS One 2013;8:e70855.

[10] McDermott S, Moran R, Platt T, Wood H, Isaac T, Dasari S. Prevalence of epilepsy in adults with mental retardation and related disabilities in primary care. American Journal on Mental Retardation 2005;110:48–56.

[11] Cooper S-A, Hughes-McCormack L, Greenlaw N, McConnachie A, Allan L, Baltzer M, et al. Management and prevalence of long-term conditions in primary health care for adults with intellectual disabilities compared with the general population: A population-based cohort study. Journal of Applied Research in Intellectual Disabilities 2018;31:68–81.

[12] Truesdale M, Melville C, Barlow F, Dunn K, Henderson A, Hughes-McCormack LA, et al. Respiratory-associated deaths in people with intellectual disabilities: a systematic review and meta-analysis. BMJ Open 2021;11:e043658.

[13] Deb S, Perera B, Krysta K, Ozer M, Bertelli M, Novell R, et al. The European guideline on the assessment and diagnosis of psychiatric disorders in adults with intellectual disabilities. Eur J Psychiatry 2022;36:11–25.

[14] Abakasanga E, Kousovista R, Cosma G, Jun GT, Kiani R, Gangadharan S. Identifying clusters on multiple long-term conditions for adults with learning disabilities. International Conference on AI in Healthcare, 2024, p. 45–58.

[15] Abakasanga E, Kousovista R, Cosma G, Jun GT, Kiani R, Gangadharan S. Cluster and trajectory analysis of multiple long-term conditions in adults with learning disabilities. International Conference on AI in Healthcare, 2024, p. 3–16.

[16] Shabnam S, Kousovista R, Abakasanga E, Kaur N, Cosma G, Akbari A, et al. Data preparation and epidemiological plan - DECODE 2024. https://doi.org/10.17605/OSF.IO/KT5FY.

[17] García-Domínguez L, Navas P, Verdugo MÁ, Arias VB. Chronic Health Conditions in Aging Individuals with Intellectual Disabilities. Int J Environ Res Public Health 2020;17.

[18] Dolan E, Lane J, Hillis G, Delanty N. Changing trends in life expectancy in intellectual disability over time. Ir Med J 2021;112.

[19] Shankar R, Rowe C, Van Hoorn A, Henley W, Laugharne R, Cox D, et al. Under representation of people with epilepsy and intellectual disability in research. PLoS One 2018;13:e0198261.

[20] Robertson J, Hatton C, Emerson E, Baines S. Prevalence of epilepsy among people with intellectual disabilities: a systematic review. Seizure 2015;29:46–62.

[21] Wallace SJ. Epilepsy in cerebral palsy. Dev Med Child Neurol 2001;43:713–7.

[22] Jindai K, Nielson CM, Vorderstrasse BA, Quiñones AR. Multimorbidity and functional limitations among adults 65 or older, NHANES 2005-2012. Prev Chronic Dis 2016;13:1–11.

[23] Castello R, Caputo M. Thyroid diseases and gender. Journal of Sex-and Gender-Specific Medicine 2019;5:136–41.

[24] Levi M, Simonetti M, Marconi E, Brignoli O, Cancian M, Masotti A, et al. Gender differences in determinants of iron-deficiency anemia: a population-based study conducted in four European countries. Ann Hematol 2019;98:1573–82.

[25] Navas P, Llorente S, Garcia L, Tassé MJ, Havercamp SM. Improving healthcare access for older adults with intellectual disability: What are the needs? Journal of Applied Research in Intellectual Disabilities 2019;32:1453–64.

[26] Bigby C. Ageing with a lifelong disability: A guide to practice, program, and policy issues for human services professionals. Jessica Kingsley Publishers; 2004.

[27] Robertson J, Emerson E, Baines S, Hatton C. Obesity and health behaviours of British adults with self-reported intellectual impairments: cross sectional survey. BMC Public Health 2014;14:1–7.

[28] Masenga SK, Kirabo A. Hypertensive heart disease: risk factors, complications and mechanisms. Front Cardiovasc Med 2023;10:1205475.

[29] De Bhailis ÁM, Kalra PA. Hypertension and the kidneys. Br J Hosp Med 2022;83:1–11.

**Supplementary Appendix**

**Comorbidity Patterns and Temporal Associations of Multiple Long-Term Conditions in Adults with Intellectual Disability: an observational study in the United Kingdom**

Georgina Cosma, Emeka Abakasanga, Rania Kousovista, Sharmin Shabnam, Navjot Kaur, Ashley Akbari, Reza Kiani, Francesco Zaccardi, Gyuchan Thomas Jun, Satheesh Gangadharan

Professor Georgina Cosma PhD, School of Science, Loughborough University, Loughborough, UK.
Dr Emeka Abakasanga PhD, School of Science, Loughborough University, Loughborough, UK.
Dr Rania Kousovista PhD, School of Science, Loughborough University, Loughborough, UK.
Dr Sharmin Shabnam PhD, Leicester Real World Evidence Unit, Leicester General Hospital, UK.
Dr Navjot Kaur, Leicester Real World Evidence Unit, Leicester General Hospital, UK.
Professor Ashley Akbari PhD, Faculty of Medicine, Health and Life Science, Swansea University, Swansea, UK.
Dr Reza Kiani MD, PhD, FRCPsych, Consultant Neuropsychiatrist and Honorary Associate Professor, Leicestershire Partnership NHS Trust & University of Leicester, UK.
Dr Francesco Zaccardi PhD, Leicester Real World Evidence Unit, Leicester General Hospital, UK.
Professor Gyuchan Thomas Jun PhD, School of Design and Creative Arts, Loughborough University, Loughborough, UK.
Professor Satheesh Gangadharan, Consultant Psychiatrist, Leicestershire Partnership NHS Trust, UK.

*Corresponding author: Professor G Cosma PhD; School of Science, Loughborough University, Loughborough, UK. Email: g.cosma@lboro.ac.uk. Phone: +44 (0)1509 564 158

**Contents**

## Section 1. Heatmaps to visualise the co-occurrence and statistical associations between conditions

**Constructing the heatmaps.** Heatmaps were constructed to visualise the co-occurrence and statistical associations between 38 conditions in male patients, and 40 conditions in female patients. For each pair of conditions, we analysed their co-occurrence patterns. We categorised patients into four mutually exclusive groups: those with both conditions, those with only the first condition, those with only the second condition, and those with neither condition. This allowed us to construct a standard 2x2 contingency table for each condition pair. To assess the statistical significance of these associations, we employed either Fisher's exact test or Pearson's chi-squared test, depending on the expected cell frequencies. Specifically, we used Fisher's exact test when any cell in the 2x2 table had an expected count less than 5; otherwise, we applied the chi-squared test. To account for multiple comparisons, we applied False Discovery Rate (FDR) correction to the p-values. We then calculated Cramer's V as a measure of effect size for each significant association. The heatmap displays FDR-corrected p-values using a blue colour gradient, with darker hues indicating stronger associations (lower p-values). We masked cells white if they had non-significant p-values ($\geq 0.05$) or small effect sizes (Cramer's V $<0.1$) to emphasise meaningful associations. The co-occurrence frequencies are displayed within each cell, with frequencies below 10 represented by a dash ("-") for privacy considerations.

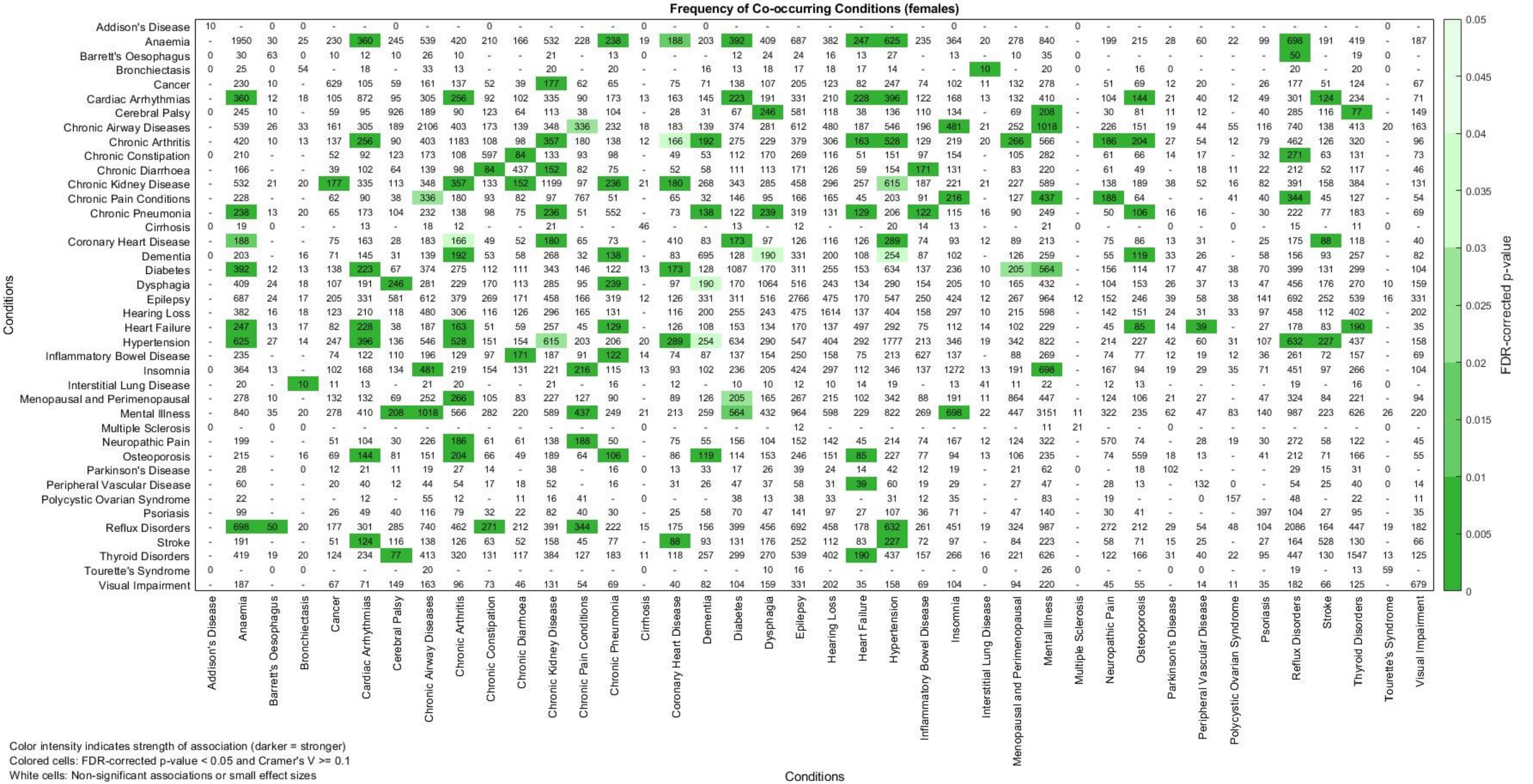
Frequency of Co-occurring Conditions (females)
Conditions
Addison's Disease
Anaemia
Barrett's Oesophagus
Bronchiectasis
Cancer
Cardiac Arrhythmias
Cerebral Palsy
Chronic Airway Diseases
Chronic Arthritis
Chronic Constipation
Chronic Diarrhoea
Chronic Kidney Disease
Chronic Pain Conditions
Chronic Pneumonia
Cirrhosis
Coronary Heart Disease
Dementia
Diabetes
Dysphagia
Epilepsy
Hearing Loss
Heart Failure
Hypertension
Inflammatory Bowel Disease
Insomnia
Interstitial Lung Disease
Menopausal and Perimenopausal
Mental Illness
Multiple Sclerosis
Neuropathic Pain
Osteoporosis
Parkinson's Disease
Peripheral Vascular Disease
Polycystic Ovarian Syndrome
Psoriasis
Reflux Disorders
Stroke
Thyroid Disorders
Tourette's Syndrome
Visual Impairment
FDR-corrected p-value
0.05
0.045
0.04
0.035
0.03
0.025
0.02
0.015
0.01
0.005
0
Color intensity indicates strength of association (darker = stronger)
Colored cells: FDR-corrected p-value < 0.05 and Cramer's V >= 0.1
White cells: Non-significant associations or small effect sizes

(a)

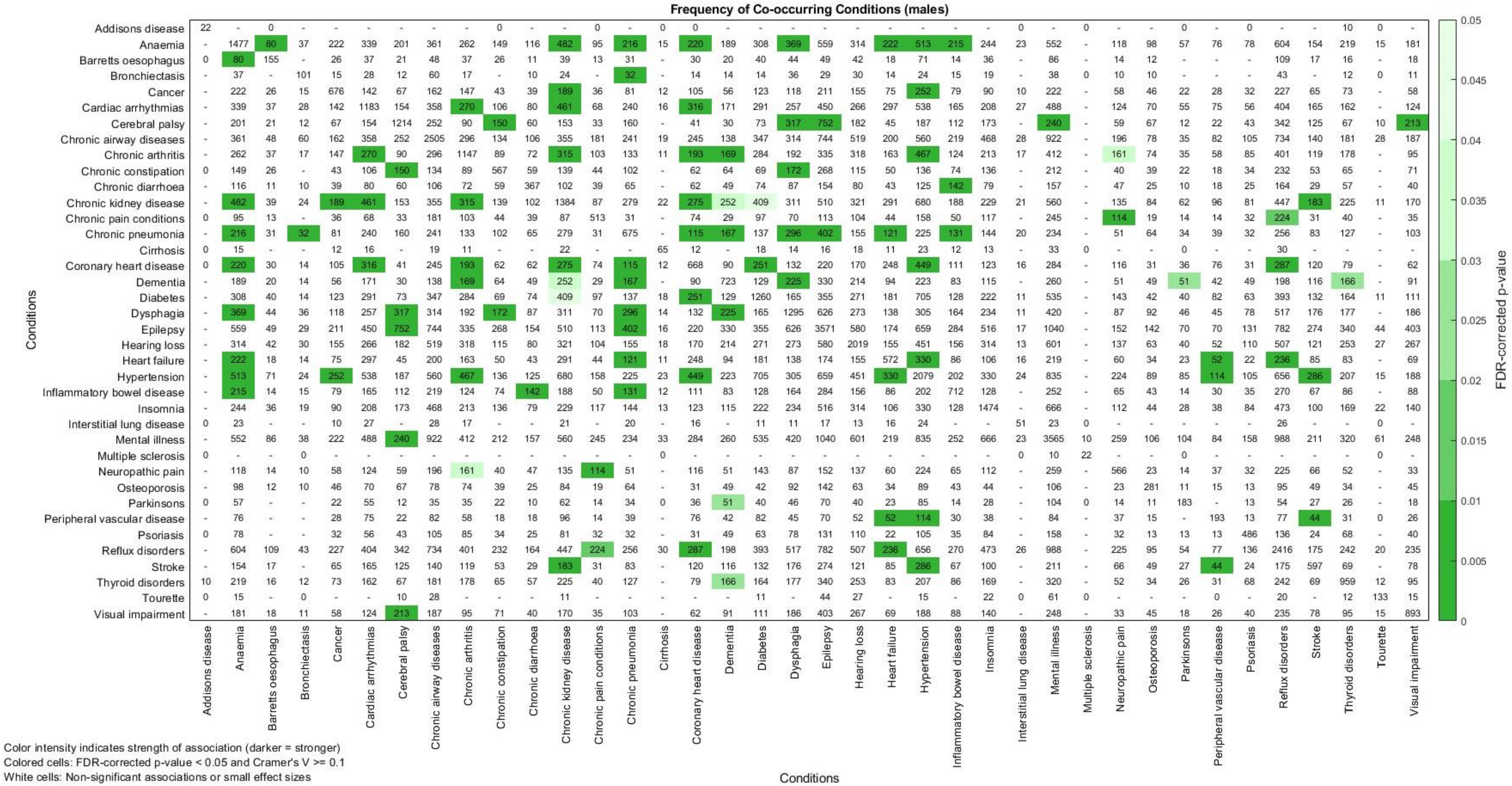

(b)

**Figure S.1:** Heatmaps of co-occurring conditions in (a) females and (b) males. Green colour intensity indicates strength of association (darker = stronger, FDR-corrected p < 0.05 and Cramer's V ≥ 0.1). White cells represent non-significant associations or small effect sizes. Numbers show co-occurrence frequency; "-" denotes frequencies between 1 and 9. The heatmap includes 40 conditions for females and 38 for males.

**Section 2. Age at diagnosis by condition for females and males**

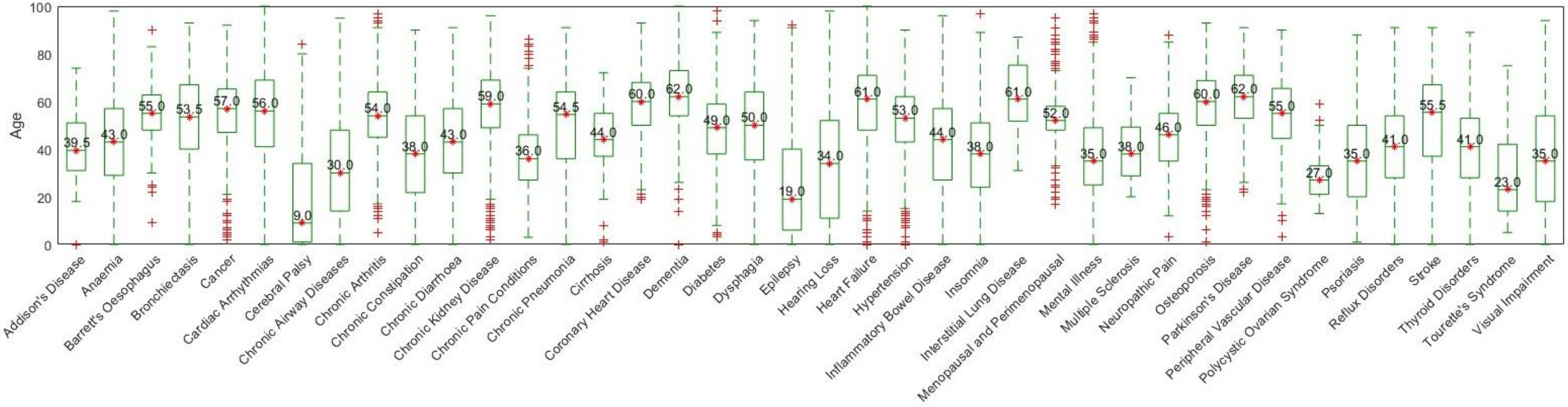

(a) Females: Age at diagnosis by condition

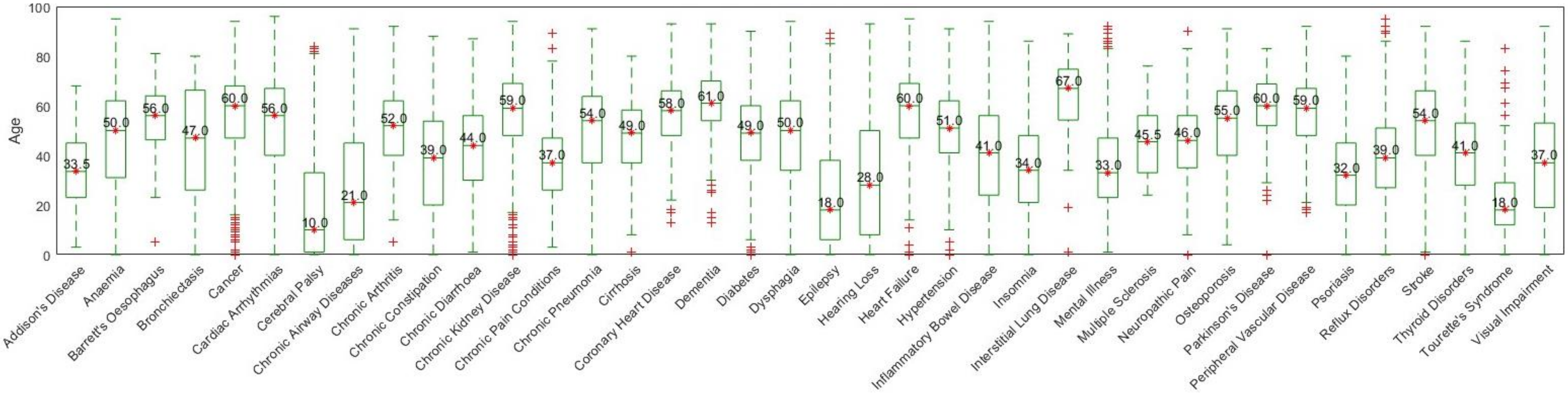

**(b)** Males: Age at diagnosis by condition

**Figure S.1: Box plots of age at diagnosis of (a) females and (b) males**

**Section 3. Long-term conditions used in the study and statistics**

**Table S.0. Long-term conditions used in the study and their system categorisation**

| **Condition** | **Included Conditions** | **Condition Category based on the system affected** |
|---|---|---|
| Addisons Disease | | Endocrine |
| Anaemia | | Blood |
| Barretts Oesophagus | | Digestive |
| Bronchiectasis | | Respiratory |
| Cancer | | Neoplasms |
| Arrhythmias | Atrial Fibrillation, Atrial Flutter, Supraventricular Tachycardia, Heart Block | Circulatory |
| Cerebral Palsy | | Nervous |
| Chronic Airway Diseases | Eosinophil Bronchitis, Chronic Obstructive Pulmonary Disease (COPD), Asthma | Respiratory |
| Chronic Arthritis | Osteoarthritis, Arthritis, Autoimmune Arthritis | Musculoskeletal |
| Chronic Constipation | | Digestive |
| Chronic Diarrhoea | | Digestive |
| Chronic Kidney Disease | | Genitourinary |
| Chronic Pain Conditions | Fibromyalgia, Pelvic Pain, Chronic Back Pain | Musculoskeletal |
| Chronic Pneumonia | | Respiratory |
| Cirrhosis | | Digestive |
| Coronary Heart Disease | | Circulatory |
| Dementia | | Mental |
| Diabetes | | Endocrine |
| Dysphagia | | Digestive |
| Epilepsy | | Nervous |
| Heart Failure | | Circulatory |
| Hearing Loss | | Ear |
| Hypertension | | Circulatory |
| Inflammatory Bowel Disease | | Digestive |
| Insomnia | | Nervous |
| Interstitial Lung Disease | Interstitial Pulmonary Fibrosis | Respiratory |
| Mental Illness | Schizophrenia, Severe Depression | Mental |
| Menopausal And Perimenopausal | | Genitourinary |
| Multiple Sclerosis | | Nervous |
| Neuropathic Pain | | Nervous |
| Osteoporosis | | Musculoskeletal |
| Parkinsons | | Nervous |
| Peripheral Vascular Disease | Peripheral Arterial Disease | Circulatory |
| Polycystic Ovary Syndrome | | Endocrine |
| Psoriasis | | Skin |
| Reflux Disorders | Gastro-oesophageal reflux disease (GORD), Gastric and Peptic Ulcer Diseases, Helicobacter Pylori, Dyspepsia | Digestive |
| Stroke | | Nervous |
| Thyroid Disorders | Hyperthyroidism, Hypoparathyroidism, All Hypo and Hyper, Hashimoto | Endocrine |
| Tourette | | Mental |
| Visual Impairment | | Eye |

**Table S.1: List of long-term conditions and statistics. For each condition, the table reports the mean age (standard deviation) and median age (IQR range) at first diagnosis, number of patients, and percentage of patients with condition, separately for males and females.**

| | Male (n=10168) | | | | Female (n=7967) | | | |
|---|---|---|---|---|---|---|---|---|
| | **Mean age at diagnosis [SD]** | **Median age at diagnosis [Q1-Q3]** | **No. of patients** | **% of patients** | **Mean age at diagnosis [SD]** | **Median age at diagnosis [Q1-Q3]** | **No. of patients** | **% of patients** |
| Addison's Disease | 34.9 [18.3] | 33.5 [23.0-45] | 22 | 0.2 | 40 [21.1] | 39.5 [31-51] | 10 | 0.1 |
| Anaemia | 46.1 [21.5] | 50 [31-62] | 1477 | 14.5 | 43.3 [19.2] | 43 [29-57] | 1950 | 24.4 |
| Barrett's Oesophagus | 54.5 [13.2] | 56 [46.2-64] | 155 | 1.5 | 54.7 [16] | 55 [48-62.8] | 63 | 0.8 |
| Bronchiectasis | 45.4 [23.2] | 47 [26-66.2] | 101 | 1 | 49.8 [22.3] | 53.5 [40-67] | 54 | 0.7 |
| Cancer | 55.8 [17.8] | 60 [47-68] | 676 | 6.6 | 55.2 [16.4] | 57 [47-65.2] | 629 | 7.9 |
| Cardiac Arrhythmias | 53.2 [18.3] | 56 [40-67] | 1183 | 11.6 | 54.2 [19.6] | 56 [41-69] | 872 | 10.9 |
| Cerebral Palsy | 18.3 [20.4] | 10 [1-33] | 1214 | 11.9 | 18.8 [21.1] | 9 [1-34] | 926 | 11.6 |
| Chronic Airway Diseases | 26.9 [22.9] | 21 [6-45] | 2505 | 24.6 | 32 [21.3] | 30 [14-48] | 2106 | 26.4 |
| Chronic Arthritis | 51.3 [14.6] | 52 [40-62] | 1147 | 11.3 | 54.1 [14.3] | 54 [45-64] | 1183 | 14.8 |
| Chronic Constipation | 37.6 [20.9] | 39 [20-53.8] | 567 | 5.6 | 37.4 [20.4] | 38 [21.8-54] | 597 | 7.5 |
| Chronic Diarrhoea | 42.5 [18.8] | 44 [30-56] | 367 | 3.6 | 43.4 [18.5] | 43 [30-57] | 437 | 5.5 |
| Chronic Kidney Disease | 57.3 [16.4] | 59 [48-69] | 1384 | 13.6 | 58.2 [16.1] | 59 [49-69] | 1199 | 15 |
| Chronic Pain Conditions | 37.9 [14.6] | 37 [26-47] | 513 | 5 | 37 [13.7] | 36 [27-46] | 767 | 9.6 |
| Chronic Pneumonia | 48.8 [21.4] | 54 [37-63.8] | 675 | 6.6 | 49.4 [21.1] | 54.5 [36-64] | 552 | 6.9 |
| Cirrhosis | 47.4 [14.6] | 49 [37-58.2] | 65 | 0.6 | 43.6 [16] | 44 [37-55] | 46 | 0.6 |
| Coronary Heart Disease | 57.1 [13.4] | 58 [48-66] | 668 | 6.6 | 58.7 [14.8] | 60 [50-68] | 410 | 5.1 |
| Dementia | 62 [12] | 61 [54-70] | 723 | 7.1 | 63.4 [13.2] | 62 [54-73] | 695 | 8.7 |
| Diabetes | 48.4 [16.1] | 49 [38-60] | 1260 | 12.4 | 48.2 [16.2] | 49 [38-59] | 1087 | 13.6 |
| Dysphagia | 47.5 [18.5] | 50 [34-62] | 1295 | 12.7 | 49 [18.9] | 50 [35.5-64] | 1064 | 13.3 |
| Epilepsy | 23.5 [20] | 18 [6-38] | 3571 | 35.1 | 24.5 [20.8] | 19 [6-40] | 2766 | 34.7 |
| Hearing Loss | 30.6 [22.9] | 28 [8-50] | 2019 | 19.9 | 34.2 [23.5] | 34 [11-52] | 1614 | 20.2 |
| Heart Failure | 56.4 [18.8] | 60 [47-69] | 572 | 5.6 | 58 [19.1] | 61 [48-71] | 497 | 6.2 |
| Hypertension | 51.2 [14.5] | 51 [41-62] | 2079 | 20.4 | 52.3 [14.8] | 53 [43-62] | 1777 | 22.3 |
| Inflammatory Bowel Disease | 40.1 [21.3] | 41 [24-56] | 712 | 7 | 42.5 [20.7] | 44 [27-57] | 627 | 7.9 |
| Insomnia | 34.7 [17.9] | 34 [21-48] | 1474 | 14.5 | 37.8 [17.7] | 38 [24-51] | 1272 | 15.9 |
| Interstitial Lung Disease | 62.6 [16.8] | 67 [54.2-74.8] | 51 | 0.5 | 61.9 [14.4] | 61 [51.8-75.2] | 41 | 0.5 |
| Menopausal and Perimenopausal | NA | NA | NA | NA | 52.8 [9.4] | 52 [48-58] | 864 | 10.8 |
| Mental Illness | 35.7 [16.3] | 33 [23-47] | 3565 | 35.1 | 37.9 [16.4] | 35 [25-49] | 3151 | 39.5 |
| Multiple Sclerosis | 46 [14.8] | 45.5 [33-56] | 22 | 0.2 | 39.4 [14.1] | 38 [28.8-49.2] | 21 | 0.3 |
| Neuropathic Pain | 46 [15.4] | 46 [35-56] | 566 | 5.6 | 45.9 [14.8] | 46 [35-55] | 570 | 7.1 |
| Osteoporosis | 52.3 [18] | 55 [40-66] | 281 | 2.8 | 57.7 [16.4] | 60 [50-68.8] | 559 | 7 |
| Parkinson's Disease | 58.4 [14] | 60 [52-68.8] | 183 | 1.8 | 60.5 [14.9] | 62 [53-71] | 102 | 1.3 |
| Peripheral Vascular Disease | 56.5 [15.5] | 59 [48-67] | 193 | 1.9 | 53.6 [17.3] | 55 [44.5-65.5] | 132 | 1.7 |
| Polycystic Ovarian Syndrome | NA | NA | NA | NA | 27.8 [8.6] | 27 [21-33] | 157 | 2 |
| Psoriasis | 33.4 [16.6] | 32 [20-45] | 486 | 4.8 | 35.5 [18.8] | 35 [20-50] | 397 | 5 |
| Reflux Disorders | 39.5 [17.1] | 39 [27-51] | 2416 | 23.8 | 41 [17.7] | 41 [28-54] | 2086 | 26.2 |
| Stroke | 48.9 [22.8] | 54 [40-66] | 597 | 5.9 | 49.8 [24.1] | 55.5 [37-67] | 528 | 6.6 |
| Thyroid Disorders | 40.6 [18.3] | 41 [28-53] | 959 | 9.4 | 40.8 [17.7] | 41 [28-53] | 1547 | 19.4 |
| Tourette's Syndrome | 23.1 [15.9] | 18 [12-29] | 133 | 1.3 | 27.6 [17.6] | 23 [14-42] | 59 | 0.7 |
| Visual Impairment | 36 [21.6] | 37 [19-53] | 893 | 8.8 | 36.3 [22.7] | 35 [18-54] | 679 | 8.5 |

**Section 4. Pair information, duration and precedence statistics**

In Tables S.2-S.7:

- Condition A: Specifies the first condition in the pair being analysed.
- Condition B: Specifies the second condition in the pair being analysed.
- Pair Freq.: Indicates the count of how often these two conditions occur together in a specific group.
- Group %: Represents the percentage of cases where this pair occurs within a specific group.
- Odds Ratio [CI]: Shows the strength of association between the two conditions, with a confidence interval.
- Median Duration Years [IQR]: Shows the median length of time the conditions co-occur, with the interquartile range.
- Precedence: Specifies which condition typically appears first.
- Directionality Freq.: Directionality frequency shows how many times the typical order of appearance is observed.
- Directionality %: Directionality percentage is the percentage of cases where the typical order of appearance is observed.

**Table S.2: Females under 45 (n=6397). Pairs with frequencies greater than or equal to 100**

| Condition pair | | Pair information and duration | | | | Pair precedence & directionality | | |
|---|---|---|---|---|---|---|---|---|
| **Condition A** | **Condition B** | **Odds Ratio [CI]** | **Pair freq.** | **Group %** | **Median duration years [IQR]** | **Precedence** | **Direct. freq.** | **Direct. %** |
| Cerebral Palsy | Epilepsy | 5.29[4.53-6.18] | 484 | 7.57 | 6.41 [1.39-15.53] | Cerebral Palsy precedes Epilepsy | 235 | 48.55 |
| Chronic Airway Diseases | Reflux Disorders | 2.60[2.27-2.98] | 402 | 6.28 | 7.78 [2.91-15.65] | Chronic Airway Diseases precedes Reflux Disorders | 282 | 70.15 |
| Insomnia | Mental Illness | 3.09[2.66-3.58] | 401 | 6.27 | 4.52 [1.78-9.13] | Mental Illness precedes Insomnia | 238 | 59.35 |
| Chronic Airway Diseases | Insomnia | 2.80[2.39-3.27] | 288 | 4.50 | 6.54 [2.97-12.20] | Chronic Airway Diseases precedes Insomnia | 196 | 68.06 |
| Chronic Pain Conditions | Mental Illness | 3.07[2.58-3.66] | 284 | 4.44 | 4.69 [1.84-8.52] | Mental Illness precedes Chronic Pain Conditions | 152 | 53.52 |
| Anaemia | Reflux Disorders | 2.39[2.05-2.79] | 284 | 4.44 | 3.74 [1.06-8.00] | Anaemia precedes Reflux Disorders | 135 | 47.54 |
| Insomnia | Reflux Disorders | 2.70[2.29-3.19] | 239 | 3.74 | 4.30 [1.51-8.24] | Insomnia precedes Reflux Disorders | 121 | 50.63 |
| Chronic Airway Diseases | Chronic Pain Conditions | 3.17[2.64-3.79] | 220 | 3.44 | 7.70 [2.90-14.02] | Chronic Airway Diseases precedes Chronic Pain Conditions | 155 | 70.45 |
| Dysphagia | Epilepsy | 2.93[2.40-3.57] | 214 | 3.35 | 13.35 [6.24-23.01] | Epilepsy precedes Dysphagia | 193 | 90.19 |
| Epilepsy | Visual Impairment | 2.87[2.35-3.50] | 209 | 3.27 | 8.04 [2.33-18.00] | Epilepsy precedes Visual Impairment | 143 | 68.42 |
| Chronic Pain Conditions | Reflux Disorders | 3.63[3.02-4.36] | 203 | 3.17 | 3.48 [1.50-6.23] | Chronic Pain Conditions precedes Reflux Disorders | 115 | 56.65 |
| Dysphagia | Reflux Disorders | 4.39[3.57-5.39] | 173 | 2.70 | 3.76 [0.83-8.13] | Reflux Disorders precedes Dysphagia | 102 | 58.96 |
| Diabetes | Hypertension | 9.73[7.76-12.19] | 144 | 2.25 | 4.86 [1.59-8.56] | Diabetes precedes Hypertension | 72 | 50.00 |
| Chronic Constipation | Reflux Disorders | 3.99[3.20-4.97] | 144 | 2.25 | 3.70 [1.53-8.35] | Chronic Constipation precedes Reflux Disorders | 85 | 59.03 |
| Mental Illness | Neuropathic Pain | 3.14[2.46-4.02] | 140 | 2.19 | 4.52 [1.62-9.07] | Mental Illness precedes Neuropathic Pain | 94 | 67.14 |
| Cerebral Palsy | Dysphagia | 5.35[4.30-6.66] | 139 | 2.17 | 14.20 [5.25-22.75] | Cerebral Palsy precedes Dysphagia | 116 | 83.45 |
| Chronic Pain Conditions | Insomnia | 3.27[2.65-4.03] | 136 | 2.13 | 3.35 [1.22-6.72] | Chronic Pain Conditions precedes Insomnia | 78 | 57.35 |
| Diabetes | Reflux Disorders | 2.48[2.00-3.09] | 127 | 1.99 | 4.75 [1.84-10.07] | Reflux Disorders precedes Diabetes | 72 | 56.69 |
| Anaemia | Dysphagia | 2.89[2.31-3.60] | 121 | 1.89 | 4.91 [1.22-9.88] | Anaemia precedes Dysphagia | 74 | 61.16 |
| Cerebral Palsy | Visual Impairment | 4.17[3.32-5.24] | 118 | 1.84 | 8.20 [2.72-15.00] | Cerebral Palsy precedes Visual Impairment | 83 | 70.34 |
| Inflammatory Bowel Disease | Reflux Disorders | 3.43[2.71-4.35] | 116 | 1.81 | 4.84 [1.89-8.09] | Reflux Disorders precedes Inflammatory Bowel Disease | 63 | 54.31 |
| Chronic Diarrhoea | Reflux Disorders | 5.34[4.10-6.96] | 110 | 1.72 | 2.97 [1.11-6.75] | Chronic Diarrhoea precedes Reflux Disorders | 55 | 50.00 |
| Chronic Pneumonia | Epilepsy | 4.58[3.36-6.24] | 110 | 1.72 | 11.32 [3.38-21.89] | Epilepsy precedes Chronic Pneumonia | 80 | 72.73 |
| Neuropathic Pain | Reflux Disorders | 3.76[2.91-4.85] | 103 | 1.61 | 4.34 [2.13-5.97] | Reflux Disorders precedes Neuropathic Pain | 75 | 72.82 |
| Chronic Pain Conditions | Neuropathic Pain | 9.54[7.31-12.44] | 100 | 1.56 | 4.48 [1.86-7.47] | Chronic Pain Conditions precedes Neuropathic Pain | 72 | 72.00 |

**Table S.3: Females between 45-64 (n=3494). Pairs with frequencies greater than or equal to 100**

| Condition pair | | Pair information and duration | | | | Pair precedence & directionality | | |
|---|---|---|---|---|---|---|---|---|
| **Condition A** | **Condition B** | **Odds Ratio [CI]** | **Pair freq.** | **Group %** | **Median duration years [IQR]** | **Precedence** | **Direct. freq.** | **Direct. %** |
| Diabetes | Hypertension | 9.39[7.74-11.40] | 236 | 6.75 | 3.52 [1.12-6.50] | Diabetes precedes Hypertension | 119 | 50.42 |
| Hypertension | Reflux Disorders | 4.44[3.70-5.32] | 212 | 6.07 | 3.51 [1.19-6.85] | Hypertension precedes Reflux Disorders | 104 | 49.06 |
| Chronic Kidney Disease | Hypertension | 4.89[4.04-5.91] | 194 | 5.55 | 3.25 [0.97-7.20] | Hypertension precedes Chronic Kidney Disease | 118 | 60.82 |
| Mental Illness | Reflux Disorders | 4.72[3.91-5.70] | 192 | 5.50 | 3.42 [1.22-6.59] | Mental Illness precedes Reflux Disorders | 106 | 55.21 |
| Chronic Arthritis | Hypertension | 4.02[3.34-4.85] | 191 | 5.47 | 4.33 [1.52-8.02] | Hypertension precedes Chronic Arthritis | 95 | 49.74 |
| Hypertension | Mental Illness | 3.02[2.52-3.61] | 191 | 5.47 | 3.83 [1.45-7.69] | Mental Illness precedes Hypertension | 108 | 56.54 |
| Hypertension | Menopausal and Perimenopausal | 3.31[2.74-3.99] | 177 | 5.07 | 3.42 [0.87-6.92] | Menopausal and Perimenopausal precedes Hypertension | 91 | 51.41 |
| Chronic Arthritis | Reflux Disorders | 5.32[4.37-6.48] | 175 | 5.01 | 3.81 [1.28-6.69] | Chronic Arthritis precedes Reflux Disorders | 88 | 50.29 |
| Menopausal and Perimenopausal | Mental Illness | 3.78[3.12-4.59] | 167 | 4.78 | 4.18 [1.47-7.91] | Mental Illness precedes Menopausal and Perimenopausal | 82 | 49.10 |
| Anaemia | Hypertension | 3.49[2.87-4.23] | 167 | 4.78 | 4.65 [1.55-7.07] | Hypertension precedes Anaemia | 90 | 53.89 |
| Chronic Airway Diseases | Hypertension | 4.74[3.86-5.83] | 158 | 4.52 | 3.34 [0.98-7.49] | Chronic Airway Diseases precedes Hypertension | 74 | 46.84 |
| Anaemia | Reflux Disorders | 4.63[3.78-5.68] | 154 | 4.41 | 1.58 [0.28-4.78] | Reflux Disorders precedes Anaemia | 67 | 43.51 |
| Chronic Arthritis | Menopausal and Perimenopausal | 4.34[3.55-5.32] | 154 | 4.41 | 3.75 [1.65-7.42] | Menopausal and Perimenopausal precedes Chronic Arthritis | 92 | 59.74 |
| Dementia | Epilepsy | 16.83[13.27-21.34] | 152 | 4.35 | 1.46 [0.34-3.26] | Dementia precedes Epilepsy | 77 | 50.66 |
| Anaemia | Chronic Kidney Disease | 5.70[4.63-7.03] | 152 | 4.35 | 2.50 [0.62-5.81] | Chronic kidney disease precedes Anaemia | 72 | 47.37 |
| Menopausal and Perimenopausal | Reflux Disorders | 3.93[3.21-4.82] | 152 | 4.35 | 4.09 [1.60-7.07] | Menopausal and Perimenopausal precedes Reflux Disorders | 84 | 55.26 |
| Chronic Arthritis | Mental Illness | 3.46[2.84-4.23] | 152 | 4.35 | 3.24 [1.24-7.94] | Mental Illness precedes Chronic Arthritis | 86 | 56.58 |
| Hypertension | Thyroid Disorders | 3.62[2.95-4.43] | 151 | 4.32 | 4.49 [1.34-6.76] | Thyroid Disorders precedes Hypertension | 85 | 56.29 |
| Anaemia | Mental Illness | 3.25[2.65-4.00] | 139 | 3.98 | 3.41 [0.62-7.01] | Mental Illness precedes Anaemia | 77 | 55.40 |
| Dysphagia | Reflux Disorders | 6.81[5.45-8.52] | 138 | 3.95 | 1.48 [0.08-4.21] | Dysphagia precedes Reflux Disorders | 64 | 46.38 |
| Chronic Kidney Disease | Mental Illness | 3.46[2.81-4.26] | 138 | 3.95 | 3.27 [0.87-8.93] | Mental Illness precedes Chronic Kidney Disease | 83 | 60.14 |
| Chronic Airway Diseases | Reflux Disorders | 5.44[4.37-6.78] | 135 | 3.86 | 3.87 [1.39-7.73] | Chronic Airway Diseases precedes Reflux Disorders | 72 | 53.33 |
| Chronic Airway Diseases | Chronic Arthritis | 5.72[4.59-7.13] | 133 | 3.81 | 4.31 [1.43-7.12] | Chronic Airway Diseases precedes Chronic Arthritis | 75 | 56.39 |
| Anaemia | Chronic Arthritis | 3.85[3.11-4.77] | 131 | 3.75 | 3.00 [0.73-6.34] | Chronic Arthritis precedes Anaemia | 69 | 52.67 |
| Hypertension | Insomnia | 4.29[3.43-5.36] | 129 | 3.69 | 3.50 [1.33-6.92] | Hypertension precedes Insomnia | 66 | 51.16 |
| Chronic Airway Diseases | Mental Illness | 4.20[3.38-5.23] | 129 | 3.69 | 3.79 [1.07-7.78] | Chronic Airway Diseases precedes Mental Illness | 61 | 47.29 |
| Chronic Kidney Disease | Reflux Disorders | 3.61[2.91-4.48] | 126 | 3.61 | 3.09 [1.60-5.91] | Reflux Disorders precedes Chronic Kidney Disease | 86 | 68.25 |
| Diabetes | Mental Illness | 3.58[2.88-4.46] | 124 | 3.55 | 4.20 [1.38-9.42] | Mental Illness precedes Diabetes | 63 | 50.81 |
| Insomnia | Mental Illness | 4.84[3.85-6.09] | 122 | 3.49 | 2.92 [0.91-7.24] | Mental Illness precedes Insomnia | 73 | 59.84 |
| Diabetes | Reflux Disorders | 4.19[3.35-5.23] | 122 | 3.49 | 4.18 [1.50-7.48] | Diabetes precedes Reflux Disorders | 60 | 49.18 |
| Chronic Arthritis | Chronic Kidney Disease | 3.34[2.67-4.18] | 114 | 3.26 | 3.43 [1.12-7.07] | Chronic Arthritis precedes Chronic Kidney Disease | 76 | 66.67 |
| Insomnia | Reflux Disorders | 5.11[4.04-6.46] | 113 | 3.23 | 2.75 [0.92-5.37] | Reflux Disorders precedes Insomnia | 58 | 51.33 |
| Chronic Kidney Disease | Thyroid Disorders | 4.25[3.38-5.34] | 113 | 3.23 | 3.05 [1.05-7.50] | Thyroid Disorders precedes Chronic Kidney Disease | 76 | 67.26 |
| Mental Illness | Thyroid Disorders | 2.87[2.30-3.59] | 113 | 3.23 | 4.40 [1.77-8.24] | Thyroid Disorders precedes Mental Illness | 56 | 49.56 |
| Anaemia | Menopausal and Perimenopausal | 2.87[2.29-3.58] | 112 | 3.21 | 3.54 [1.71-6.82] | Menopausal and Perimenopausal precedes Anaemia | 69 | 61.61 |
| Coronary Heart Disease | Hypertension | 11.49[8.58-15.38] | 111 | 3.18 | 2.17 [0.36-6.38] | Hypertension precedes Coronary Heart Disease | 65 | 58.56 |
| Epilepsy | Mental Illness | 3.17[2.52-3.98] | 110 | 3.15 | 3.68 [0.79-8.01] | Mental Illness precedes Epilepsy | 51 | 46.36 |
| Hearing Loss | Hypertension | 3.15[2.49-3.97] | 107 | 3.06 | 3.51 [1.44-7.34] | Hypertension precedes Hearing Loss | 51 | 47.66 |
| Cardiac Arrhythmias | Hypertension | 4.40[3.44-5.62] | 106 | 3.03 | 2.93 [0.69-7.55] | Hypertension precedes Cardiac Arrhythmias | 55 | 51.89 |
| Epilepsy | Thyroid Disorders | 4.86[3.83-6.17] | 105 | 3.01 | 3.73 [1.13-6.73] | Thyroid Disorders precedes Epilepsy | 52 | 49.52 |
| Chronic Airway Diseases | Menopausal and Perimenopausal | 3.65[2.89-4.62] | 104 | 2.98 | 3.72 [1.35-7.30] | Chronic Airway Diseases precedes Menopausal and Perimenopausal | 52 | 50.00 |
| Reflux Disorders | Thyroid Disorders | 3.04[2.41-3.83] | 104 | 2.98 | 4.34 [2.06-6.93] | Thyroid Disorders precedes Reflux Disorders | 60 | 57.69 |
| Chronic Kidney Disease | Diabetes | 3.94[3.11-4.99] | 102 | 2.92 | 4.26 [1.60-8.40] | Diabetes precedes Chronic Kidney Disease | 73 | 71.57 |
| Anaemia | Dysphagia | 4.70[3.68-5.99] | 101 | 2.89 | 2.45 [0.11-5.23] | Anaemia precedes Dysphagia | 42 | 41.58 |
| Chronic Arthritis | Diabetes | 3.38[2.67-4.28] | 101 | 2.89 | 3.65 [1.70-7.31] | No clear precedence | 101 | 100.00 |
| Chronic Arthritis | Thyroid Disorders | 3.13[2.47-3.95] | 101 | 2.89 | 4.77 [1.93-8.34] | Thyroid Disorders precedes Chronic Arthritis | 55 | 54.46 |
| Menopausal and Perimenopausal | Thyroid Disorders | 2.95[2.34-3.73] | 101 | 2.89 | 4.42 [1.09-7.33] | Thyroid Disorders precedes Menopausal and Perimenopausal | 57 | 56.44 |

**Table S.4: Females 65+ (n=1292). Pairs with frequencies greater or equal to 70**

| Condition pair | | Pair information and duration | | | | Pair precedence & directionality | | |
|---|---|---|---|---|---|---|---|---|
| **Condition A** | **Condition B** | **Odds Ratio [CI]** | **Pair freq.** | **Group %** | **Median duration years [IQR]** | **Precedence** | **Direct. freq.** | **Direct. %** |
| Chronic Kidney Disease | Hypertension | 20.00[15.71-25.47] | 153 | 11.84 | 4.09 [1.35-9.30] | Hypertension precedes Chronic Kidney Disease | 102 | 66.67 |
| Anaemia | Chronic Kidney Disease | 23.80[18.44-30.72] | 144 | 11.15 | 2.88 [0.47-5.81] | Anaemia precedes Chronic Kidney Disease | 76 | 52.78 |
| Cardiac Arrhythmias | Chronic Kidney Disease | 22.77[17.64-29.39] | 141 | 10.91 | 2.18 [0.10-5.92] | Cardiac Arrhythmias precedes Chronic Kidney Disease | 64 | 45.39 |
| Chronic Kidney Disease | Dementia | 19.75[15.30-25.50] | 133 | 10.29 | 2.03 [0.53-5.56] | Dementia precedes Chronic Kidney Disease | 65 | 48.87 |
| Chronic Kidney Disease | Heart Failure | 32.76[24.35-44.09] | 120 | 9.29 | 1.99 [0.17-4.78] | Heart Failure precedes Chronic Kidney Disease | 56 | 46.67 |
| Cardiac Arrhythmias | Hypertension | 18.91[14.50-24.65] | 115 | 8.90 | 3.69 [1.52-9.02] | Hypertension precedes Cardiac Arrhythmias | 80 | 69.57 |
| Chronic Arthritis | Hypertension | 20.71[15.79-27.14] | 113 | 8.75 | 4.00 [1.43-7.31] | Hypertension precedes Chronic Arthritis | 55 | 48.67 |
| Chronic Arthritis | Chronic Kidney Disease | 14.34[10.99-18.71] | 107 | 8.28 | 4.14 [1.81-7.86] | Chronic Arthritis precedes Chronic Kidney Disease | 71 | 66.36 |
| Cardiac Arrhythmias | Heart Failure | 39.96[29.42-54.28] | 106 | 8.20 | 0.31 [0.00-3.98] | Cardiac Arrhythmias precedes Heart Failure | 47 | 44.34 |
| Anaemia | Hypertension | 15.93[12.19-20.82] | 106 | 8.20 | 3.11 [0.93-8.71] | Hypertension precedes Anaemia | 64 | 60.38 |
| Dementia | Hypertension | 15.63[11.96-20.44] | 105 | 8.13 | 4.55 [1.30-10.29] | Hypertension precedes Dementia | 79 | 75.24 |
| Anaemia | Cardiac Arrhythmias | 19.16[14.52-25.27] | 102 | 7.89 | 3.78 [0.85-7.92] | Anaemia precedes Cardiac Arrhythmias | 62 | 60.78 |
| Chronic Kidney Disease | Mental Illness | 15.83[11.91-21.05] | 95 | 7.35 | 2.23 [0.20-6.42] | Mental Illness precedes Chronic Kidney Disease | 47 | 49.47 |
| Chronic Kidney Disease | Dysphagia | [11.02-19.39] | 94 | 7.28 | 2.15 [0.27-5.98] | Chronic kidney disease precedes Dysphagia | 50 | 53.19 |
| Anaemia | Chronic Arthritis | 17.54[13.19-23.32] | 92 | 7.12 | 2.66 [1.23-5.28] | Chronic Arthritis precedes Anaemia | 48 | 52.17 |
| Cardiac Arrhythmias | Dementia | 15.65[11.81-20.72] | 92 | 7.12 | 3.55 [0.87-5.90] | Cardiac Arrhythmias precedes Dementia | 46 | 50.00 |
| Chronic Kidney Disease | Reflux Disorders | 16.49[12.25-22.20] | 88 | 6.81 | 2.69 [0.48-6.85] | Reflux Disorders precedes Chronic Kidney Disease | 58 | 65.91 |
| Chronic Kidney Disease | Osteoporosis | 15.88[11.79-21.40] | 86 | 6.66 | 4.05 [1.67-6.19] | Osteoporosis precedes Chronic Kidney Disease | 51 | 59.30 |
| Chronic Arthritis | Dementia | 15.10[11.32-20.15] | 85 | 6.58 | 5.59 [1.62-9.43] | Chronic Arthritis precedes Dementia | 62 | 72.94 |
| Heart Failure | Hypertension | 17.52[12.95-23.72] | 82 | 6.35 | 4.14 [0.80-9.99] | Hypertension precedes Heart Failure | 57 | 69.51 |
| Anaemia | Heart Failure | 21.08[15.47-28.71] | 80 | 6.19 | 2.76 [0.82-5.55] | Anaemia precedes Heart Failure | 44 | 55.00 |
| Anaemia | Dementia | 12.10[9.07-16.14] | 80 | 6.19 | 2.97 [0.98-5.89] | Anaemia precedes Dementia | 50 | 62.50 |
| Chronic Kidney Disease | Diabetes | 18.87[13.64-26.11] | 77 | 5.96 | 4.32 [1.17-8.41] | Diabetes precedes Chronic Kidney Disease | 49 | 63.64 |
| Cardiac Arrhythmias | Chronic Arthritis | 12.72[9.48-17.07] | 77 | 5.96 | 6.19 [3.11-11.08] | Chronic Arthritis precedes Cardiac Arrhythmias | 60 | 77.92 |
| Dementia | Mental Illness | 16.13[11.88-21.91] | 75 | 5.80 | 2.05 [0.39-6.77] | Mental Illness precedes Dementia | 46 | 61.33 |
| Dementia | Dysphagia | 15.32[11.31-20.76] | 75 | 5.80 | 0.96 [0.23-3.96] | Dementia precedes Dysphagia | 35 | 46.67 |
| Chronic Kidney Disease | Hearing Loss | 13.08[9.62-17.78] | 75 | 5.80 | 3.34 [0.86-7.09] | Hearing Loss precedes Chronic Kidney Disease | 49 | 65.33 |
| Diabetes | Hypertension | 21.19[15.23-29.47] | 73 | 5.65 | 3.72 [1.39-6.74] | Diabetes precedes Hypertension | 35 | 47.95 |
| Hypertension | Osteoporosis | 14.64[10.74-19.95] | 73 | 5.65 | 3.34 [0.85-6.83] | Hypertension precedes Osteoporosis | 44 | 60.27 |
| Hearing Loss | Hypertension | 15.21[11.12-20.82] | 72 | 5.57 | 4.00 [1.43-8.94] | Hearing Loss precedes Hypertension | 41 | 56.94 |
| Anaemia | Mental Illness | 14.99[11.01-20.40] | 72 | 5.57 | 4.15 [1.36-8.66] | Anaemia precedes Mental Illness | 38 | 52.78 |
| Hypertension | Mental Illness | 11.96[8.84-16.20] | 72 | 5.57 | 3.87 [1.20-7.85] | Hypertension precedes Mental Illness | 41 | 56.94 |
| Chronic Kidney Disease | Coronary Heart Disease | 19.03[13.53-26.75] | 70 | 5.42 | 2.93 [0.71-6.51] | Coronary Heart Disease precedes Chronic Kidney Disease | 43 | 61.43 |
| Dysphagia | Hypertension | 10.84[8.01-14.68] | 70 | 5.42 | 2.35 [0.66-6.15] | Hypertension precedes Dysphagia | 44 | 62.86 |

**Table S.5: Males under 45 (n=8296). Pairs with frequencies greater or equal to 50**

| Condition pair | | Pair information and duration | | | | Pair precedence & directionality | | |
|---|---|---|---|---|---|---|---|---|
| Condition A | Condition B | Odds Ratio [CI] | Pair freq. | Group % | Median duration years [IQR] | Precedence | Direct. Freq. | Direct. % |
| Cerebral Palsy | Epilepsy | 4.27[3.74-4.87] | 614 | 7.40 | 6.67 [1.26-17.90] | Cerebral Palsy precedes Epilepsy | 271 | 44.14 |
| Dysphagia | Epilepsy | 2.42[2.03-2.88] | 255 | 3.07 | 16.03 [6.85-22.81] | Epilepsy precedes Dysphagia | 224 | 87.84 |
| Dysphagia | Reflux Disorders | 4.47[3.72-5.36] | 216 | 2.60 | 3.55 [0.48-9.50] | Reflux Disorders precedes Dysphagia | 121 | 56.02 |
| Cerebral Palsy | Dysphagia | 6.26[5.18-7.57] | 200 | 2.41 | 16.70 [5.68-24.57] | Cerebral Palsy precedes Dysphagia | 176 | 88.00 |
| Anaemia | Reflux Disorders | 3.13[2.61-3.74] | 197 | 2.37 | 4.27 [0.41-11.05] | Reflux Disorders precedes Anaemia | 84 | 42.64 |
| Cerebral Palsy | Visual Impairment | 4.44[3.66-5.40] | 168 | 2.03 | 10.45 [2.36-18.22] | Cerebral Palsy precedes Visual Impairment | 117 | 69.64 |
| Chronic Constipation | Epilepsy | 2.51[2.02-3.13] | 164 | 1.98 | 12.56 [5.62-19.69] | Epilepsy precedes Chronic Constipation | 124 | 75.61 |
| Diabetes | Hypertension | 8.51[6.92-10.48] | 160 | 1.93 | 3.67 [0.89-7.61] | Diabetes precedes Hypertension | 83 | 51.88 |
| Chronic Pneumonia | Epilepsy | 4.10[3.12-5.39] | 136 | 1.64 | 12.78 [3.51-22.10] | Epilepsy precedes Chronic Pneumonia | 99 | 72.79 |
| Inflammatory Bowel Disease | Reflux Disorders | 3.13[2.52-3.89] | 131 | 1.58 | 3.84 [1.41-9.54] | Reflux Disorders precedes Inflammatory Bowel Disease | 70 | 53.44 |
| Chronic Pain Conditions | Reflux Disorders | 2.99[2.37-3.76] | 114 | 1.37 | 3.73 [1.64-8.26] | Chronic Pain Conditions precedes Reflux Disorders | 74 | 64.91 |
| Anaemia | Dysphagia | 4.99[3.97-6.26] | 112 | 1.35 | 2.94 [0.00-8.67] | Anaemia precedes Dysphagia | 49 | 43.75 |
| Cerebral Palsy | Chronic Constipation | 4.78[3.77-6.06] | 111 | 1.34 | 9.12 [3.12-17.47] | Cerebral Palsy precedes Chronic Constipation | 77 | 69.37 |
| Chronic Constipation | Reflux Disorders | 2.93[2.32-3.72] | 108 | 1.30 | 4.43 [1.69-10.15] | Chronic Constipation precedes Reflux Disorders | 67 | 62.04 |
| Cardiac Arrhythmias | Reflux Disorders | 2.57[2.03-3.25] | 106 | 1.28 | 4.48 [1.02-10.47] | Reflux Disorders precedes Cardiac Arrhythmias | 70 | 66.04 |
| Cerebral Palsy | Chronic Pneumonia | 6.96[5.29-9.16] | 94 | 1.13 | 8.86 [2.97-17.93] | Cerebral Palsy precedes Chronic Pneumonia | 72 | 76.60 |
| Chronic Pneumonia | Dysphagia | 14.02 [10.54-18.65] | 88 | 1.06 | 4.05 [1.10-9.67] | Chronic Pneumonia precedes Dysphagia | 64 | 72.73 |
| Cerebral Palsy | Stroke | 7.87[5.87-10.56] | 86 | 1.04 | 0.43 [0.00-11.80] | Cerebral Palsy precedes Stroke | 31 | 36.05 |
| Chronic Pneumonia | Reflux Disorders | 3.84[2.92-5.07] | 86 | 1.04 | 5.75 [1.72-10.33] | Chronic Pneumonia precedes Reflux Disorders | 44 | 51.16 |
| Dysphagia | Visual Impairment | 3.78[2.94-4.85] | 85 | 1.02 | 7.46 [1.66-14.99] | Visual Impairment precedes Dysphagia | 61 | 71.76 |
| Chronic Kidney Disease | Hypertension | 6.44[4.89-8.49] | 78 | 0.94 | 2.90 [0.11-6.79] | Hypertension precedes Chronic Kidney Disease | 45 | 57.69 |
| Chronic Constipation | Insomnia | 2.75[2.11-3.59] | 75 | 0.90 | 3.80 [0.99-9.87] | Chronic Constipation precedes Insomnia | 46 | 61.33 |
| Chronic Constipation | Dysphagia | 5.72[4.34-7.54] | 73 | 0.88 | 8.47 [5.16-15.07] | Chronic Constipation precedes Dysphagia | 65 | 89.04 |
| Anaemia | Inflammatory Bowel Disease | 4.04[3.09-5.29] | 73 | 0.88 | 2.89 [0.49-8.18] | Inflammatory Bowel Disease precedes Anaemia | 44 | 60.27 |
| Chronic Diarrhoea | Reflux Disorders | 3.79[2.80-5.13] | 71 | 0.86 | 3.32 [1.47-8.93] | Reflux Disorders precedes Chronic Diarrhoea | 37 | 52.11 |
| Chronic Diarrhoea | Inflammatory Bowel Disease | 18.06[13.14-24.84] | 69 | 0.83 | 0.06 [0.00-0.83] | Chronic Diarrhoea precedes Inflammatory Bowel Disease | 33 | 47.83 |
| Anaemia | Chronic Kidney Disease | 5.64[4.22-7.55] | 66 | 0.80 | 4.25 [1.31-8.80] | Chronic kidney disease precedes Anaemia | 33 | 50.00 |
| Cardiac Arrhythmias | Hypertension | 3.44[2.59-4.55] | 65 | 0.78 | 2.81 [0.84-10.73] | Hypertension precedes Cardiac Arrhythmias | 32 | 49.23 |
| Cardiac Arrhythmias | Chronic Kidney Disease | 9.89[7.29-13.42] | 63 | 0.76 | 2.00 [0.02-7.19] | Cardiac Arrhythmias precedes Chronic Kidney Disease | 29 | 46.03 |
| Cerebral Palsy | Chronic Kidney Disease | 2.70[2.02-3.61] | 62 | 0.75 | 15.72 [3.86-24.01] | Cerebral Palsy precedes Chronic Kidney Disease | 53 | 85.48 |
| Anaemia | Cardiac Arrhythmias | 3.55[2.66-4.73] | 61 | 0.74 | 4.43 [1.12-10.35] | Anaemia precedes Cardiac Arrhythmias | 34 | 55.74 |
| Cardiac Arrhythmias | Dysphagia | 3.90[2.91-5.23] | 59 | 0.71 | 3.34 [0.56-8.85] | Cardiac Arrhythmias precedes Dysphagia | 28 | 47.46 |
| Dysphagia | Inflammatory Bowel Disease | 3.35[2.49-4.50] | 57 | 0.69 | 4.79 [2.07-8.34] | Inflammatory Bowel Disease precedes Dysphagia | 42 | 73.68 |
| Chronic Pneumonia | Insomnia | 3.24[2.38-4.41] | 57 | 0.69 | 8.47 [3.61-16.30] | Chronic Pneumonia precedes Insomnia | 34 | 59.65 |
| Chronic Arthritis | Hypertension | 2.77[2.06-3.71] | 57 | 0.69 | 3.90 [1.99-7.84] | Hypertension precedes Chronic Arthritis | 28 | 49.12 |
| Chronic Kidney Disease | Diabetes | 5.37[3.94-7.33] | 55 | 0.66 | 4.21 [1.04-12.53] | Diabetes precedes Chronic Kidney Disease | 38 | 69.09 |
| Chronic Arthritis | Thyroid Disorders | 3.30[2.43-4.46] | 54 | 0.65 | 6.67 [2.70-10.86] | Thyroid Disorders precedes Chronic Arthritis | 38 | 70.37 |

**Table S.6: Males 45-64 (n=3969). Pairs with frequencies greater or equal to 100**

| Condition pair | | Pair information and duration | | | | Pair precedence & directionality | | |
|---|---|---|---|---|---|---|---|---|
| **Condition A** | **Condition B** | **Odds Ratio [CI]** | **Pair freq.** | **Group %** | **Median duration years [IQR]** | **Precedence** | **Direct. Freq.** | **Direct. %** |
| Diabetes | Hypertension | 9.73[8.10-11.68] | 253 | 6.37 | 3.41 [0.86-6.69] | Hypertension precedes Diabetes | 118 | 46.64 |
| Hypertension | Mental Illness | 4.03[3.39-4.78] | 220 | 5.54 | 4.70 [1.61-8.12] | Mental Illness precedes Hypertension | 113 | 51.36 |
| Hypertension | Reflux Disorders | 4.55[3.82-5.41] | 219 | 5.52 | 4.26 [1.61-7.48] | Reflux Disorders precedes Hypertension | 115 | 52.51 |
| Chronic Kidney Disease | Hypertension | 5.36[4.46-6.43] | 206 | 5.19 | 3.64 [0.65-7.48] | Hypertension precedes Chronic Kidney Disease | 138 | 66.99 |
| Anaemia | Hypertension | 4.53[3.75-5.48] | 178 | 4.48 | 3.95 [1.19-6.97] | Hypertension precedes Anaemia | 118 | 66.29 |
| Anaemia | Reflux Disorders | 7.24[5.95-8.81] | 177 | 4.46 | 1.84 [0.24-5.12] | Reflux Disorders precedes Anaemia | 81 | 45.76 |
| Anaemia | Chronic Kidney Disease | 9.12[7.45-11.16] | 175 | 4.41 | 2.60 [0.76-5.34] | Anaemia precedes Chronic Kidney Disease | 92 | 52.57 |
| Coronary Heart Disease | Hypertension | 9.40[7.56-11.68] | 173 | 4.36 | 2.59 [0.43-6.10] | Hypertension precedes Coronary Heart Disease | 84 | 48.55 |
| Chronic Arthritis | Hypertension | 4.34[3.58-5.27] | 171 | 4.31 | 3.68 [1.25-7.44] | Hypertension precedes Chronic Arthritis | 90 | 52.63 |
| Mental Illness | Reflux Disorders | 4.30[3.56-5.20] | 170 | 4.28 | 2.79 [0.71-6.17] | Mental Illness precedes Reflux Disorders | 96 | 56.47 |
| Cardiac Arrhythmias | Hypertension | 5.52[4.50-6.77] | 160 | 4.03 | 2.66 [0.60-6.48] | Hypertension precedes Cardiac Arrhythmias | 100 | 62.50 |
| Dementia | Epilepsy | 15.48[12.35-19.40] | 155 | 3.91 | 2.24 [0.51-4.53] | Dementia precedes Epilepsy | 83 | 53.55 |
| Chronic Airway Diseases | Hypertension | 5.09[4.13-6.26] | 152 | 3.83 | 4.56 [0.88-7.74] | Hypertension precedes Chronic Airway Diseases | 82 | 53.95 |
| Cardiac Arrhythmias | Chronic Kidney Disease | 9.23[7.42-11.48] | 144 | 3.63 | 0.90 [0.03-4.21] | Cardiac Arrhythmias precedes Chronic Kidney Disease | 70 | 48.61 |
| Dysphagia | Reflux Disorders | 5.95[4.82-7.36] | 140 | 3.53 | 2.01 [0.14-4.93] | Reflux Disorders precedes Dysphagia | 76 | 54.29 |
| Chronic Airway Diseases | Reflux Disorders | 6.02[4.83-7.51] | 128 | 3.22 | 3.12 [1.61-7.80] | Reflux Disorders precedes Chronic Airway Diseases | 70 | 54.69 |
| Chronic Kidney Disease | Diabetes | 5.71[4.59-7.10] | 127 | 3.20 | 3.52 [0.73-7.23] | Diabetes precedes Chronic Kidney Disease | 84 | 66.14 |
| Chronic Kidney Disease | Mental Illness | 3.46[2.80-4.27] | 126 | 3.17 | 1.96 [0.55-5.42] | Mental Illness precedes Chronic Kidney Disease | 82 | 65.08 |
| Hearing Loss | Hypertension | 3.14[2.54-3.89] | 124 | 3.12 | 4.73 [2.01-7.65] | Hearing Loss precedes Hypertension | 69 | 55.65 |
| Chronic Kidney Disease | Reflux Disorders | 3.73[3.01-4.62] | 123 | 3.10 | 3.60 [0.79-7.60] | Reflux Disorders precedes Chronic Kidney Disease | 78 | 63.41 |
| Anaemia | Mental Illness | 3.59[2.90-4.45] | 123 | 3.10 | 3.68 [1.26-7.87] | Mental Illness precedes Anaemia | 81 | 65.85 |
| Cardiac Arrhythmias | Coronary Heart Disease | 13.56[10.61-17.32] | 118 | 2.97 | 1.11 [0.01-5.31] | Coronary Heart Disease precedes Cardiac Arrhythmias | 51 | 43.22 |
| Chronic Arthritis | Mental Illness | 3.41[2.74-4.24] | 117 | 2.95 | 2.79 [0.74-6.83] | Mental Illness precedes Chronic Arthritis | 57 | 48.72 |
| Dementia | Dysphagia | 9.29[7.31-11.82] | 112 | 2.82 | 2.36 [0.69-3.85] | Dementia precedes Dysphagia | 68 | 60.71 |
| Hypertension | Insomnia | 3.87[3.07-4.86] | 112 | 2.82 | 3.46 [1.06-6.50] | Insomnia precedes Hypertension | 55 | 49.11 |
| Chronic Kidney Disease | Dysphagia | 5.12[4.07-6.44] | 110 | 2.77 | 2.03 [0.30-5.10] | Dysphagia precedes Chronic Kidney Disease | 58 | 52.73 |
| Chronic Arthritis | Reflux Disorders | 3.44[2.75-4.30] | 109 | 2.75 | 4.01 [1.97-7.62] | Reflux Disorders precedes Chronic Arthritis | 62 | 56.88 |
| Diabetes | Reflux Disorders | 3.52[2.81-4.42] | 107 | 2.70 | 4.21 [1.52-7.16] | Reflux Disorders precedes Diabetes | 64 | 59.81 |
| Diabetes | Mental Illness | 3.16[2.53-3.96] | 107 | 2.70 | 4.51 [1.00-7.77] | Mental Illness precedes Diabetes | 63 | 58.88 |
| Chronic Pneumonia | Dysphagia | 13.10[10.13-16.95] | 106 | 2.67 | 1.71 [0.20-4.61] | Chronic Pneumonia precedes Dysphagia | 59 | 55.66 |
| Chronic Kidney Disease | Heart Failure | 14.93[11.38-19.60] | 105 | 2.65 | 1.52 [0.04-4.21] | Heart Failure precedes Chronic Kidney Disease | 51 | 48.57 |
| Insomnia | Reflux Disorders | 5.41[4.27-6.87] | 105 | 2.65 | 2.85 [1.23-6.25] | Insomnia precedes Reflux Disorders | 52 | 49.52 |
| Insomnia | Mental Illness | 4.86[3.84-6.16] | 105 | 2.65 | 3.15 [1.00-6.60] | Mental Illness precedes Insomnia | 59 | 56.19 |
| Cardiac Arrhythmias | Heart Failure | 21.20[16.05-28.01] | 103 | 2.60 | 0.86 [0.02-4.19] | Cardiac Arrhythmias precedes Heart Failure | 54 | 52.43 |
| Chronic Arthritis | Chronic Kidney Disease | 3.96[3.14-4.99] | 103 | 2.60 | 2.26 [0.75-5.93] | Chronic Arthritis precedes Chronic Kidney Disease | 61 | 59.22 |
| Coronary Heart Disease | Heart Failure | 27.51[20.67-36.59] | 101 | 2.54 | 0.78 [0.01-3.46] | Coronary Heart Disease precedes Heart Failure | 50 | 49.50 |
| Anaemia | Dysphagia | 4.84[3.82-6.13] | 101 | 2.54 | 2.05 [0.19-5.21] | Anaemia precedes Dysphagia | 46 | 45.54 |
| Anaemia | Diabetes | 4.34[3.43-5.49] | 101 | 2.54 | 3.09 [1.19-7.33] | Diabetes precedes Anaemia | 64 | 63.37 |
| Chronic Kidney Disease | Coronary Heart Disease | 6.89[5.38-8.82] | 100 | 2.52 | 1.53 [0.28-5.34] | Coronary Heart Disease precedes Chronic Kidney Disease | 55 | 55.00 |
| Hypertension | Stroke | 6.72[5.16-8.76] | 100 | 2.52 | 2.90 [0.67-5.78] | Hypertension precedes Stroke | 58 | 58.00 |

**Table S.7: Males 65+ (n=1413). Pairs with minimum prevalence (i.e. group percentage) greater or equal to 5%**

| Condition pair | | Pair information and duration | | | | Pair precedence & directionality | | |
|---|---|---|---|---|---|---|---|---|
| **Condition A** | **Condition B** | **Odds Ratio [CI]** | **Pair freq.** | **Group %** | **Median duration years [IQR]** | **Precedence** | **Direct. freq.** | **Direct. %** |
| Chronic Kidney Disease | Hypertension | 26.19[20.82-32.95] | 180 | 12.74 | 3.12 [0.72-6.83] | Hypertension precedes Chronic Kidney Disease | 125 | 69.44 |
| Cardiac Arrhythmias | Chronic Kidney Disease | 27.06[21.47-34.10] | 179 | 12.67 | 1.12 [0.04-4.86] | Cardiac Arrhythmias precedes Chronic Kidney Disease | 84 | 46.93 |
| Anaemia | Chronic Kidney Disease | 26.34[20.51-33.84] | 148 | 10.47 | 2.13 [0.29-4.43] | Anaemia precedes Chronic Kidney Disease | 67 | 45.27 |
| Cardiac Arrhythmias | Hypertension | 19.62[15.32-25.13] | 128 | 9.06 | 3.15 [0.54-7.00] | Hypertension precedes Cardiac Arrhythmias | 74 | 57.81 |
| Chronic Kidney Disease | Heart Failure | 33.41[25.04-44.57] | 123 | 8.70 | 2.09 [0.10-4.62] | Heart Failure precedes Chronic Kidney Disease | 58 | 47.15 |
| Cardiac Arrhythmias | Heart Failure | 39.99[29.81-53.65] | 112 | 7.93 | 0.60 [0.01-3.92] | Cardiac Arrhythmias precedes Heart Failure | 60 | 53.57 |
| Anaemia | Cardiac Arrhythmias | 21.26[16.31-27.71] | 110 | 7.78 | 2.11 [0.36-5.15] | Anaemia precedes Cardiac Arrhythmias | 57 | 51.82 |
| Chronic Kidney Disease | Dementia | 14.52[11.19-18.84] | 107 | 7.57 | 2.53 [0.77-5.16] | Chronic kidney disease precedes Dementia | 57 | 53.27 |
| Chronic Kidney Disease | Diabetes | 23.06[17.28-30.77] | 103 | 7.29 | 3.68 [0.99-6.87] | Diabetes precedes Chronic Kidney Disease | 65 | 63.11 |
| Cardiac Arrhythmias | Coronary Heart Disease | 34.42[25.39-46.67] | 97 | 6.86 | 0.87 [0.00-5.08] | Cardiac Arrhythmias precedes Coronary Heart Disease | 41 | 42.27 |
| Chronic Kidney Disease | Coronary Heart Disease | 23.15[17.20-31.15] | 97 | 6.86 | 2.28 [0.31-5.37] | Coronary Heart Disease precedes Chronic Kidney Disease | 58 | 59.79 |
| Anaemia | Hypertension | 16.15[12.34-21.14] | 97 | 6.86 | 4.10 [1.31-7.48] | Hypertension precedes Anaemia | 59 | 60.82 |
| Chronic Arthritis | Chronic Kidney Disease | 19.73[14.76-26.38] | 95 | 6.72 | 3.93 [1.19-7.90] | Chronic Arthritis precedes Chronic Kidney Disease | 60 | 63.16 |
| Chronic Kidney Disease | Dysphagia | 12.76[9.71-16.77] | 92 | 6.51 | 1.96 [0.17-4.65] | Chronic kidney disease precedes Dysphagia | 55 | 59.78 |
| Coronary Heart Disease | Hypertension | 28.15[20.77-38.15] | 90 | 6.37 | 2.35 [0.09-4.97] | Hypertension precedes Coronary Heart Disease | 42 | 46.67 |
| Heart Failure | Hypertension | 22.58[16.81-30.34] | 88 | 6.23 | 1.66 [0.37-6.31] | Hypertension precedes Heart Failure | 47 | 53.41 |
| Chronic Kidney Disease | Reflux Disorders | 18.64[13.84-25.10] | 88 | 6.23 | 2.50 [0.32-6.01] | Reflux Disorders precedes Chronic Kidney Disease | 54 | 61.36 |
| Cancer | Chronic Kidney Disease | 13.96[10.52-18.51] | 88 | 6.23 | 2.02 [0.40-5.12] | Cancer precedes Chronic Kidney Disease | 45 | 51.14 |
| Diabetes | Hypertension | 22.64[16.80-30.51] | 86 | 6.09 | 2.87 [0.85-6.05] | Hypertension precedes Diabetes | 39 | 45.35 |
| Cardiac Arrhythmias | Dementia | 14.52[10.99-19.20] | 86 | 6.09 | 2.50 [0.87-5.28] | Cardiac Arrhythmias precedes Dementia | 43 | 50.00 |
| Dementia | Hypertension | 13.79[10.43-18.22] | 85 | 6.02 | 4.23 [1.50-7.76] | Hypertension precedes Dementia | 58 | 68.24 |
| Coronary Heart Disease | Heart Failure | 54.34[39.03-75.66] | 83 | 5.87 | 0.90 [0.00-3.42] | Coronary Heart Disease precedes Heart Failure | 38 | 45.78 |
| Chronic Kidney Disease | Chronic Pneumonia | 26.02[18.68-36.23] | 82 | 5.80 | 1.28 [0.25-5.68] | Chronic Pneumonia precedes Chronic Kidney Disease | 53 | 64.63 |
| Chronic Airway Diseases | Chronic Kidney Disease | 16.41[12.13-22.20] | 81 | 5.73 | 2.86 [0.82-6.13] | Chronic Airway Diseases precedes Chronic Kidney Disease | 43 | 53.09 |
| Anaemia | Reflux Disorders | 30.45[22.24-41.69] | 80 | 5.66 | 1.15 [0.07-4.14] | Reflux Disorders precedes Anaemia | 38 | 47.50 |
| Cardiac Arrhythmias | Chronic Arthritis | 19.19[14.16-26.02] | 77 | 5.45 | 3.66 [0.88-7.09] | Chronic Arthritis precedes Cardiac Arrhythmias | 51 | 66.23 |
| Anaemia | Dementia | 15.83[11.79-21.25] | 76 | 5.38 | 2.52 [0.87-6.47] | Anaemia precedes Dementia | 43 | 56.58 |
| Cardiac Arrhythmias | Chronic Airway Diseases | 19.62[14.35-26.82] | 73 | 5.17 | 2.31 [0.27-7.20] | Chronic Airway Diseases precedes Cardiac Arrhythmias | 39 | 53.42 |
| Chronic Arthritis | Hypertension | 16.89[12.43-22.95] | 73 | 5.17 | 4.18 [1.88-7.24] | Hypertension precedes Chronic Arthritis | 35 | 47.95 |
| Anaemia | Cancer | 19.13[14.06-26.01] | 72 | 5.10 | 1.72 [0.15-4.57] | Anaemia precedes Cancer | 35 | 48.61 |
| Cardiac Arrhythmias | Diabetes | 16.74[12.31-22.76] | 72 | 5.10 | 3.71 [0.70-8.06] | Diabetes precedes Cardiac Arrhythmias | 39 | 54.17 |
| Chronic Kidney Disease | Mental Illness | 10.99[8.16-14.82] | 72 | 5.10 | 1.49 [0.00-4.04] | Mental Illness precedes Chronic Kidney Disease | 27 | 37.50 |

**Section 5. Sensitivity analysis**

The table presents key metrics for condition trajectory analysis across varying odds ratio thresholds.

- The OR_Threshold column indicates the minimum odds ratio required for a disease relationship to be considered significant.
- Num_Trajectories displays the count of unique disease progression pathways identified in the dataset.
- Coverage_Percent reflects the proportion of the study population whose condition patterns are represented within the identified trajectories.
- System_Pairs denotes the number of distinct condition combinations that meet the threshold criteria.
- The temporal aspects of condition progression are captured in three duration metrics: Median_Duration represents the central tendency of time between disease occurrences, whilst Q1_Duration and Q3_Duration indicate the first and third quartiles of the progression intervals, respectively. All duration measurements are expressed in years.

**Table S.8: Results of the sensitivity analysis**

| Females <45 | | | | | | |
|---|---|---|---|---|---|---|
| **OR Threshold** | **Num Trajectories** | **Coverage Percent** | **System Pairs** | **Median Duration** | **Q1 Duration** | **Q3 Duration** |
| 2 | 41 | 49.7 | 17 | 4.64 | 1.77 | 9.99 |
| 3 | 28 | 31.51 | 14 | 4.48 | 1.56 | 9.56 |
| 4 | 12 | 13.13 | 7 | 4.48 | 1.49 | 10.73 |
| 5 | 7 | 8.86 | 3 | 4.67 | 1.59 | 8.56 |
| 6 | 4 | 3.13 | 2 | 4.48 | 1.72 | 8.02 |
| 7 | 4 | 3.13 | 2 | 4.48 | 1.72 | 8.02 |
| 8 | 4 | 3.13 | 2 | 4.48 | 1.72 | 8.02 |
| 9 | 3 | 2.51 | 2 | 4.67 | 1.72 | 7.47 |
| 10 | 1 | 0.6 | 0 | 0 | 0 | 1.16 |
| 11 | 1 | 0.6 | 0 | 0 | 0 | 1.16 |
| 12 | 1 | 0.6 | 0 | 0 | 0 | 1.16 |
| 13 | 1 | 0.6 | 0 | 0 | 0 | 1.16 |
| 14 | 1 | 0.6 | 0 | 0 | 0 | 1.16 |
| 15 | 1 | 0.6 | 0 | 0 | 0 | 1.16 |
| **Females 45-64** | | | | | | |
| **OR Threshold** | **Num Trajectories** | **Coverage Percent** | **System Pairs** | **Median Duration** | **Q1 Duration** | **Q3 Duration** |
| 2 | 97 | 100 | 45 | 3.43 | 1.13 | 7.07 |
| 3 | 89 | 100 | 44 | 3.42 | 1.12 | 7.07 |
| 4 | 50 | 80.98 | 30 | 3.35 | 1.04 | 6.88 |
| 5 | 23 | 35.52 | 15 | 2.82 | 0.67 | 6.39 |
| 6 | 9 | 15.13 | 6 | 2.17 | 0.35 | 5.28 |
| 7 | 8 | 13.15 | 6 | 2.2 | 0.36 | 5.54 |
| 8 | 7 | 11.73 | 6 | 2.23 | 0.36 | 5.79 |
| 9 | 6 | 10.49 | 5 | 2.24 | 0.36 | 6.08 |
| 10 | 3 | 5.01 | 2 | 2.17 | 0.35 | 6.38 |
| 11 | 2 | 3.76 | 1 | 1.82 | 0.35 | 4.82 |
| 12 | 1 | 2.18 | 1 | 1.46 | 0.34 | 3.26 |
| 13 | 1 | 2.18 | 1 | 1.46 | 0.34 | 3.26 |
| 14 | 1 | 2.18 | 1 | 1.46 | 0.34 | 3.26 |
| 15 | 1 | 2.18 | 1 | 1.46 | 0.34 | 3.26 |
| **Females 65+** | | | | | | |
| **OR Threshold** | **Num Trajectories** | **Coverage Percent** | **System Pairs** | **Median Duration** | **Q1 Duration** | **Q3 Duration** |
| 2 | 34 | 100 | 17 | 3.34 | 0.87 | 6.8 |
| 3 | 34 | 100 | 17 | 3.34 | 0.87 | 6.8 |
| 4 | 34 | 100 | 17 | 3.34 | 0.87 | 6.8 |
| 5 | 34 | 100 | 17 | 3.34 | 0.87 | 6.8 |
| 6 | 34 | 100 | 17 | 3.34 | 0.87 | 6.8 |
| 7 | 34 | 100 | 17 | 3.34 | 0.87 | 6.8 |
| 8 | 34 | 100 | 17 | 3.34 | 0.87 | 6.8 |
| 9 | 34 | 100 | 17 | 3.34 | 0.87 | 6.8 |
| 10 | 34 | 100 | 17 | 3.34 | 0.87 | 6.8 |
| 11 | 33 | 100 | 16 | 3.34 | 0.9 | 6.83 |
| 12 | 32 | 100 | 16 | 3.34 | 0.87 | 6.8 |
| 13 | 30 | 100 | 16 | 3.34 | 0.86 | 6.8 |
| 14 | 29 | 100 | 15 | 3.34 | 0.86 | 6.77 |
| 15 | 25 | 95.98 | 14 | 3.11 | 0.86 | 6.74 |
| **Males <45** | | | | | | |
| **OR Threshold** | **Num Trajectories** | **Coverage Percent** | **System Pairs** | **Median Duration** | **Q1 Duration** | **Q3 Duration** |
| 2 | 25 | 21.7 | 10 | 4.43 | 1.64 | 10.47 |
| 3 | 19 | 16.75 | 8 | 4.27 | 1.47 | 11.05 |
| 4 | 14 | 13.31 | 7 | 5.36 | 1.81 | 13.44 |
| 5 | 7 | 4.7 | 5 | 4.05 | 2.04 | 11.8 |
| 6 | 6 | 4.26 | 5 | 3.86 | 1.1 | 10.74 |
| 7 | 3 | 2.01 | 2 | 3.67 | 1 | 9.67 |
| 8 | 2 | 1.49 | 2 | 3.86 | 1 | 8.64 |
| 9 | 1 | 0.53 | 1 | 4.05 | 1.1 | 9.67 |
| 10 | 1 | 0.53 | 1 | 4.05 | 1.1 | 9.67 |
| 11 | 1 | 0.53 | 1 | 4.05 | 1.1 | 9.67 |
| 12 | 1 | 0.53 | 1 | 4.05 | 1.1 | 9.67 |
| 13 | 1 | 0.53 | 1 | 4.05 | 1.1 | 9.67 |
| 14 | 1 | 0.53 | 1 | 4.05 | 1.1 | 9.67 |
| 15 | 0 | 0 | 0 | 0 | 0 | 0 |

| Male 45-64 | | | | | | |
|---|---|---|---|---|---|---|
| OR Threshold | Num Trajectories | Coverage Percent | System Pairs | Median Duration | Q1 Duration | Q3 Duration |
| 2 | 85 | 100 | 42 | 3.07 | 0.79 | 6.52 |
| 3 | 81 | 100 | 40 | 2.99 | 0.76 | 6.48 |
| 4 | 60 | 85.95 | 30 | 2.66 | 0.7 | 6 |
| 5 | 40 | 57.61 | 20 | 2.3 | 0.52 | 5.41 |
| 6 | 27 | 38.25 | 15 | 2.14 | 0.42 | 5.31 |
| 7 | 19 | 29 | 12 | 1.84 | 0.29 | 4.61 |
| 8 | 16 | 24.72 | 11 | 1.96 | 0.37 | 4.57 |
| 9 | 14 | 22.89 | 9 | 1.75 | 0.26 | 4.37 |
| 10 | 8 | 10.93 | 5 | 1.32 | 0.08 | 4.2 |
| 11 | 7 | 9.88 | 4 | 1.52 | 0.04 | 4.21 |
| 12 | 7 | 9.88 | 4 | 1.52 | 0.04 | 4.21 |
| 13 | 7 | 9.88 | 4 | 1.52 | 0.04 | 4.21 |
| 14 | 5 | 7.05 | 3 | 1.52 | 0.04 | 4.19 |
| 15 | 3 | 4.52 | 1 | 0.86 | 0.02 | 4.19 |

| Males 65+ | | | | | | |
|---|---|---|---|---|---|---|
| OR Threshold | Num Trajectories | Coverage Percent | System Pairs | Median Duration | Q1 Duration | Q3 Duration |
| 2 | 36 | 100 | 18 | 2.33 | 0.42 | 5.52 |
| 3 | 36 | 100 | 18 | 2.33 | 0.42 | 5.52 |
| 4 | 36 | 100 | 18 | 2.33 | 0.42 | 5.52 |
| 5 | 36 | 100 | 18 | 2.33 | 0.42 | 5.52 |
| 6 | 36 | 100 | 18 | 2.33 | 0.42 | 5.52 |
| 7 | 36 | 100 | 18 | 2.33 | 0.42 | 5.52 |
| 8 | 36 | 100 | 18 | 2.33 | 0.42 | 5.52 |
| 9 | 36 | 100 | 18 | 2.33 | 0.42 | 5.52 |
| 10 | 36 | 100 | 18 | 2.33 | 0.42 | 5.52 |
| 11 | 35 | 100 | 18 | 2.35 | 0.42 | 5.68 |
| 12 | 35 | 100 | 18 | 2.35 | 0.42 | 5.68 |
| 13 | 34 | 100 | 18 | 2.42 | 0.48 | 5.72 |
| 14 | 32 | 100 | 17 | 2.42 | 0.48 | 5.72 |
| 15 | 30 | 100 | 15 | 2.33 | 0.4 | 5.89 |

**Section 6. Prevalence of conditions in the general and intellectual disability (ID) population with MLTCs**

**Table A.9: Comparing prevalence of conditions in the general and intellectual disability (ID) population.**

| **Data from the Quality and Outcomes Framework, 2020-21** | | **Data from our study (CPRD GOLD)** | | | | |
|---|---|---|---|---|---|---|
| **General population (quality & outcomes framework 2020-2021)**<br>**Condition** | **Population prevalence**<br>**%** | **ID population (our study)**<br>**Condition or condition group** | **Included conditions** | **Prevalence in males**<br>**% males** | **Prevalence in females**<br>**% females** | **Total prevalence in CPRD**<br>**%** |
| Atrial fibrillation | 2.05 | Cardiac arrhythmias | Atrial fibrillation, atrial flutter, supraventricular tachycardia, heart block | 11.63 | 10.93 | 11.32 |
| Coronary heart disease | 3.05 | Coronary heart disease | | 6.57 | 5.14 | 5.94 |
| Heart failure | 0.91 | Heart failure | | 5.63 | 6.23 | 5.89 |
| Hypertension | 13.93 | Hypertension | | 20.45 | 22.28 | 21.25 |
| Peripheral arterial disease | 0.59 | Peripheral vascular disease | Peripheral arterial disease | 1.9 | 1.65 | 1.79 |
| Stroke and transient ischaemic attack | 1.80 | Stroke | | 5.87 | 6.62 | 6.20 |
| Asthma and COPD | 8.30 | Chronic airway diseases | Eosinophil bronchitis, chronic obstructive pulmonary disease (COPD), asthma | 24.64 | 26.4 | 25.41 |
| Cancer | 3.21 | Cancer | | 6.65 | 7.89 | 7.20 |
| Chronic kidney disease | 3.96 | Chronic kidney disease | | 13.61 | 15.03 | 14.23 |
| Diabetes mellitius | 7.11 | Diabetes | | 12.39 | 13.63 | 12.94 |
| Palliative care | 0.47 | - | | | | |
| Dementia | 0.71 | Dementia | | 7.11 | 8.71 | 7.81 |
| Depression | 12.29 | Mental illness | Schizophrenia, severe depression | 35.06 | 39.51 | 37.02 |
| Epilepsy | 0.80 | Epilepsy | | 35.12 | 34.68 | 34.93 |
| Learning disabilities | 0.53 | - | - | - | - | - |
| Mental health | 0.95 | - | - | - | - | - |
| Osteoporosis | 0.76 | Osteoporosis | | 2.76 | 7.01 | 4.63 |
| Rheumatoid arthritis | 0.77 | Chronic arthritis | Osteoarthritis, arthritis, autoimmune arthritis | 11.28 | 14.83 | 12.84 |
| Non-diabetic hyperglycaemia | 5.31 | - | - | - | - | - |
| *Source: https://digital.nhs.uk/data-and-information/publications/statistical/quality-and-outcomes-framework-achievement-prevalence-and-exceptions-data/2020-21#summary | | Source: CPRD<br>Total males: 0.56%<br>Total females: 0.44%<br>Total prevalence = (%prevalence in males * % of male population) + (%prevalence in females * % of female population) | | | | |